\documentclass[
  aps,
  prb,
  twocolumn,
  superscriptaddress,
  floatfix
]{revtex4-2}
\usepackage{float}

\usepackage[utf8]{inputenc}
\usepackage[english]{babel}
\usepackage{microtype}

\usepackage{amsmath,amssymb,amsthm}
\usepackage{bm}
\usepackage{esint}
\usepackage{stmaryrd}
\usepackage{dsfont}
\usepackage{braket}

\usepackage{graphicx}
\usepackage{xcolor}
\usepackage[normalem]{ulem}
\usepackage{cancel}

\usepackage{xcolor}
\usepackage[most]{tcolorbox}

\usepackage{hyperref}
\hypersetup{
  colorlinks=true,
  linkcolor=black,
  citecolor=blue,
  urlcolor=blue
}

\theoremstyle{plain}

\theoremstyle{definition}

\theoremstyle{remark}

\begin{document}
% --------------------------------------------------
\title{Efficient Optimization of Tensor Rings with Low-Rank Environments}

\author{Matthieu Jeannin}
\email{matt.jeannin@gmail.com}
\affiliation{Ludwig-Maximilians-Universität München, 80539 Munich, Germany}

\author{Alessandro Chessari}
\email{chessari.alessandro@gmail.com}
\affiliation{Technische Universität Wien, 1040 Vienna, Austria}

\author{Jan Von Delft}
\affiliation{Ludwig-Maximilians-Universität München, 80539 Munich, Germany}

\begin{abstract}

Tensor-ring (TR) decompositions provide a natural representation of periodic systems but are difficult to optimize because the closed geometry prevents a global canonical form and leads to costly, ill-conditioned environments. Existing periodic Density Matrix Renormalization Group (DMRG) methods alleviate this difficulty by compressing long environments to a low-rank representation, reducing local operations to $\mathcal{O}(p\chi^3)$, where $p$ is the retained environment rank. Here, we extend this approach to an efficient two-site Ring-DMRG algorithm and improve its numerical robustness through appropriate gauge transformations and a generalized Davidson solver. 
A central challenge in the two-site formulation is the truncation step, which must account for the surrounding environment. When this environment is sufficiently separable, a suitable change of frame reduces the truncation to an ordinary singular value decomposition.  When it is not separable, we instead use an alternating optimization that retains the full environment.

The computational advantage of
tensor rings over tensor trains depends on the scaling of the required environment rank $p$ with bond dimension
and system size. For critical systems, the
divergent correlation length makes long-range observables remain sensitive to the periodic geometry
as the thermodynamic limit is approached. In this regime, the environment rank required to reach a
fixed accuracy in the observable does not grow with the system size. Ring-DMRG therefore retains the same
cubic scaling with bond dimension as standard DMRG, but at a substantially smaller bond
dimension. The low-rank environment construction also provides a systematic generalization of belief propagation
(BP), with standard BP recovered in the rank-one limit.

\end{abstract}

\maketitle
\section{Introduction}

The density matrix renormalization group (DMRG), introduced by S. White in the early 1990s \cite{White92,White93,Schollwock11}, has become one of the most accurate methods for computing ground states of one-dimensional quantum many-body systems. In DMRG, the ground state is approximated by a matrix product state (MPS) \cite{Rommer97}, whose accuracy is controlled by the bond dimension \(\chi\). MPS are also known as tensor trains (TT) in numerical linear algebra and have motivated a range of DMRG-inspired construction and optimization methods, including DMRG-cross, tensor cross interpolation, and alternating optimization schemes \cite{Oseledets10,Savostyanov11,Dolgov14,Dolgov20,Chertkov22}. These methods typically retain the favorable \(\mathcal{O}(\chi^3)\) local linear-algebra complexity of standard DMRG.

\begin{figure}
	\centering
	\includegraphics[width=\linewidth]{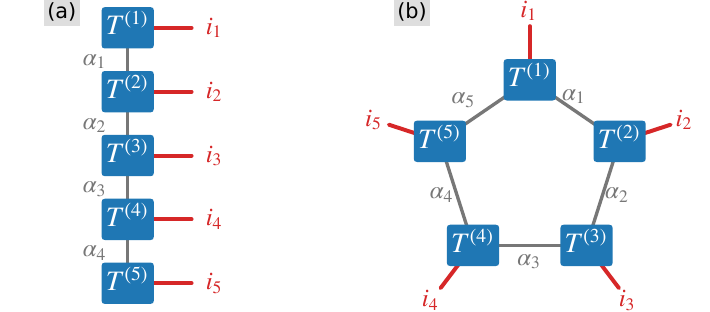}
	\caption{
		Diagrammatic representation of a tensor train or matrix product state (a) and a tensor ring or periodic matrix product state (b).
	}
	\label{fig:tt_tr}
\end{figure}

\begin{figure}[!t]
	\centerline{	\includegraphics[width=\linewidth]{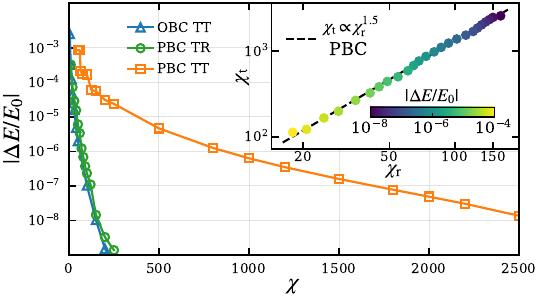} }
	\caption{
		Convergence of the ground-state energy for the spin-$1/2$ Heisenberg chain
		with $N=500$ with periodic boundary conditions. We compare standard DMRG in tensor-train form with open
		boundary conditions (OBC TT), standard DMRG with periodic boundary conditions
		(PBC TT), and Ring-DMRG with a tensor-ring ansatz (PBC TR). The relative energy error
		$|E-E_0|/|E_0|$ is shown as a function of the bond dimension $\chi$.
		For the OBC calculation, $E_0$ is the converged OBC energy, whereas
		both periodic calculations are compared with the converged PBC energy at
		$\chi_{\rm t}=3000$.
		Inset: Tensor-train bond dimension $\chi_{\rm t}$ versus tensor-ring bond
		dimension $\chi_{\rm r}$ required to reach the same relative energy error for PBC (interpolated on the data).
		The dashed line shows the fit
		$\chi_{\rm t}\propto\chi_{\rm r}^{1.5}$.
	} \label{fig:N500}
\end{figure}

For systems with essentially one-dimensional entanglement, TT provide remarkably accurate approximations at moderate bond dimension, making DMRG highly efficient in practice \cite{Jeckelmann02,Hastings07}. More generally, however, the efficiency of the representation depends on how well the underlying geometry and correlations can be encoded by the one-dimensional virtual structure \cite{Verstraete04,Eisert10}.
Periodic boundary conditions provide a particularly important example, where a natural approach is to close the virtual indices of the MPS, yielding a periodic MPS, also known as a tensor ring (TR), as illustrated in Fig.~\ref{fig} \cite{Verstraete04}.  Representing such a state using an open-boundary tensor train may require a substantially larger bond dimension, which in the worst case can grow quadratically with that of the tensor ring \cite{Gendiar00,Verstraete04,Handschuh12}. This difference is illustrated in Fig.~\ref{fig:N500}, which shows the ground-state energy error for the periodic spin-$1/2$ Heisenberg chain with system size $N=500$. The tensor-ring representation (with bond dimension $\chi_r$) is substantially more compact than the corresponding tensor train ($\chi_t$): achieving the same accuracy requires $\chi_{\mathrm{t}}\propto \chi_{\mathrm{r}}^{1.5}$. The same closed-chain tensor-network structure appears in numerical multilinear algebra as the tensor-ring decomposition \cite{Zhao16}, which has since found applications in machine learning, tensor completion, image recovery and processing, and high-dimensional data representation \cite{Wang18,Yuan18,Xu20,Sedighin21}. 

The tensor ring ansatz can represent periodic correlations much more efficiently: Fig.~\ref{fig:N500} shows the ground-state energy error for the periodic spin-$1/2$ Heisenberg chain with system size $N=500$. The tensor-ring representation (with bond dimension $\chi_r$) is substantially more compact than the corresponding tensor train ($\chi_t$): achieving the same accuracy requires $\chi_{\mathrm{t}}\propto \chi_{\mathrm{r}}^{1.5}$. At the same time, Ring-DMRG exhibits a convergence with bond dimension comparable to that of standard DMRG applied to the open-boundary ground state.

Closing a train into a tensor ring comes at a price: it removes several of the structural properties that make tensor trains easy to manipulate. First, unlike tensor trains, tensor rings do not generally admit a global canonical form in which the local norm environment becomes the identity \cite{Schollwock11}. Local optimization therefore involves nontrivial norm and Hamiltonian environments and leads naturally to generalized eigenvalue problems \cite{Moler73}. Since these environments act on doubled virtual spaces of dimension \(\chi^2\), their direct manipulation can result in computational costs as large as \(\mathcal{O}(\chi^6)\), far from the typical \(\mathcal{O}(\chi^3)\) complexity of tensor trains. Closely related bottlenecks arise in tensor ring algorithms from numerical multilinear algebra, including alternating least-squares methods combined with sampling \cite{Khoo19,Mickelin20,Yu22,Malik22_1,Malik22_2,Jeannin25_phd}, CUR-type decompositions \cite{Espig12}, Riemannian optimization \cite{Gao2026}, and, more recently, exact reconstruction schemes \cite{Chen25}.

The second issue is that the lack of a canonical form makes the parametrization redundant: the same tensor can admit different tensor-ring decompositions with different internal ranks, so the optimal rank cannot be determined locally. Moreover, because the norm environment is non-trivial, even truncation by singular value decomposition (SVD) of a local tensor block need not be locally optimal \cite{Batselier18,Ye18,Yuan20,Sedighin21r,Qiu25}.
This has direct consequences for DMRG optimization. In a tensor train, a two-site update temporarily enlarges the bond dimension, after which a local singular-value decomposition (SVD) provides a well-controlled truncation back to the target bond dimension. In a tensor ring, by contrast, this local truncation is no longer optimal \cite{SanchezFernan25}.
 
A key route around the first bottleneck was introduced in periodic DMRG by Pippan \emph{et al.} \cite{Pippan10,Pirvu11}. The central observation is that, although individual transfer matrices are generally full rank, long products of transfer matrices often possess a rapidly decaying singular-value spectrum. The effective periodic environment can therefore be approximated by a much smaller number of dominant singular directions, reducing the cost of applying it to approximately \(\mathcal{O}(p\chi^3)\), where \(p\) is the retained environment rank. We refer to this low-rank periodic-DMRG approach as Ring-DMRG. A similar scaling is obtained in a variational method based on Monte Carlo optimization \cite{Sandvik07}. Rossini \emph{et al.} subsequently improved the numerical stability of Ring-DMRG by solving the generalized eigenvalue problem in a frame with better conditioned environments \cite{Rossini11}. Related formulations were developed for translationally invariant systems \cite{Pirvu11}, expressed in the modern tensor network picture \cite{Weyrauch13}, and extended to infinite-system calculations \cite{Dey16}. Nevertheless, Ring-DMRG has remained expensive in practice, a robust two-site update is still lacking, and the regimes in which it offers a genuine advantage over tensor-train methods have not been clearly identified.

In this work, we address these challenges. In Sec.~\ref{sec:DMRG_PBC}, we revisit DMRG with periodic boundary conditions using low-rank environments computed through randomized SVD~\cite{Halko11}. We improve the stability of the local generalized eigenvalue problem by combining a Davidson eigensolver (Sec.~\ref{sec:solvers}) with a change of frame that improves the conditioning of the norm environment (Sec.~\ref{sec:frame}). In Sec.~\ref{sec:two_site}, we introduce a two-site truncation scheme that replaces the naive local SVD by a weighted low-rank approximation. This truncation scheme yields a two-site Ring-DMRG algorithm that preserves \(U(1)\) symmetry while allowing both the bond dimension and the symmetry sectors to adapt during the optimization (Sec.~\ref{sec:two_site_dmrg}). Using the same truncation scheme and a Trotter decomposition of the Hamiltonian, we compute the real-time evolution of correlation functions within the ring topology (Sec.~\ref{sec:time_evol}).

Since Ring-DMRG scales as \(\mathcal{O}(p\chi_{\rm r}^3)\), whereas standard DMRG in MPS form scales as \(\mathcal{O}(\chi_{\rm t}^3)\), we identify in Sec.~\ref{sec:regime} the regimes in which the ring topology is advantageous. For critical systems, we find that although resolving the divergent correlation length requires increasing the bond dimension with the system size, the required environment rank $p$ remains approximately independent of \(N\), while \(\chi_{\rm t}\sim\chi_{\rm r}^{1.5}\). In this case, Ring-DMRG becomes computationally preferable.

Finally, in Sec.~\ref{sec:rsvd}, we interpret the low-rank environment construction based on randomized SVD as a generalized belief-propagation scheme that retains several dominant environment directions rather than a single message.

\section{DMRG for periodic boundary conditions}
\label{sec:DMRG_PBC}

This section reviews the successive optimization strategies proposed for finding ground states in the tensor-ring format and introduces our improvements. We refer to the resulting algorithm as Ring-DMRG.

Consider a general pure state of an \(N\)-site quantum system with local Hilbert-space dimension \(d\),
\begin{equation}
\ket{\phi}=\sum_{i_1,\ldots,i_N}\phi(i_1,\ldots,i_N)\ket{i_1,\ldots,i_N}\,,
\end{equation}
where \(i_\ell=1,\ldots,d\) labels a local basis state at site \(\ell\). The amplitudes \(\phi(i_1,\ldots,i_N)\) form an \(N\)th-order tensor. A tensor ring (TR) represents this coefficient tensor as a cyclic contraction of third-order local tensors \(T^{(\ell)}\). For fixed physical index \(i_\ell\), we denote by \(T^{(\ell)}_{i_\ell}\) the \(\chi_{\ell-1}\times\chi_\ell\) matrix with entries
\(\left[T^{(\ell)}_{i_\ell}\right]_{\alpha_{\ell-1},\alpha_\ell}\), where \(\alpha_\ell=1,\ldots,\chi_\ell\) labels the virtual bond between sites \(\ell\) and \(\ell+1\). Explicitly,
\begin{equation}
\phi(i_1,\ldots,i_N)
=
\left[T^{(1)}_{i_1}\right]_{\alpha_{N},\alpha_{1}}\!
\left[T^{(2)}_{i_2}\right]_{\alpha_{1},\alpha_{2}}
\!\!\!\!\cdots \left[T^{(N)}_{i_N}\right]_{\alpha_{N-1},\alpha_{N}}\!.
\label{eq:tensor_ring}
\end{equation}
Repeated indices are implicitly summed over according to the Einstein summation convention, which we adopt throughout the remainder of this work. When no ambiguity arises Greek indices will be left implicit and not written explicitly. Periodicity is imposed by identifying \(\alpha_0\equiv\alpha_N\). We denote by \(\chi_\ell\) the dimension of the \(\ell\)th virtual bond and define the maximal bond dimension as
\begin{equation}
\chi := \max_\ell \chi_\ell.
\end{equation}
The periodic contraction in Eq.~\eqref{eq:tensor_ring} can be written as the matrix trace
\begin{equation}
\phi(i_1,\ldots,i_N)
=
\operatorname{Tr}\!\left(
T^{(1)}_{i_1}
T^{(2)}_{i_2}
\cdots
T^{(N)}_{i_N}
\right).
\label{eq:tensor_ring_trace}
\end{equation}
The TR format is the periodic analogue of tensor trains (TTs): while a TT has open virtual boundaries (\(\chi_N=1\)), a TR closes the bond between sites \(N\) and \(1\), enabling correlations across the periodic boundary (Fig.~\ref{fig:tt_tr}). TRs are therefore well suited to ground states of systems with periodic boundary conditions \cite{Verstraete04,Pippan10}.

We use analogous notation for matrix product operators (MPOs) in ring form.
Consider a general operator acting on the \(N\)-site Hilbert space,
\begin{equation}
\mathcal H
=
\sum_{\mathbf i,\mathbf j}
H_{\mathbf i,\mathbf j}
\ket{\mathbf i}\bra{\mathbf j},
\end{equation}
where \(\mathbf i=(i_1,\ldots,i_N)\) and \(\mathbf j=(j_1,\ldots,j_N)\). We denote the fourth-order tensor $W^{(\ell)}$ with entries
$\left[W^{(\ell)}_{i_\ell,j_\ell}\right]_{\omega_{\ell-1},\omega_\ell}$, where \(i_\ell\) and \(j_\ell\) are physical indices and \(\omega_\ell\) virtual index of bond dimension \(D_\ell\). The matrix elements of the full operator are then
\begin{equation}
H_{i_1,j_1,\ldots,i_N,j_N}
=
\operatorname{Tr}\!\left(
W^{(1)}_{i_1,j_1}
W^{(2)}_{i_2,j_2}
\cdots
W^{(N)}_{i_N,j_N}
\right).
\label{eq:ring_mpo}
\end{equation}
Further details of the periodic MPO construction for the spin-\(1/2\) Heisenberg Hamiltonian are given in App.~\ref{app:ring_mpo}.

\begin{figure}[!b]
	\centering
	\includegraphics[width=\linewidth]{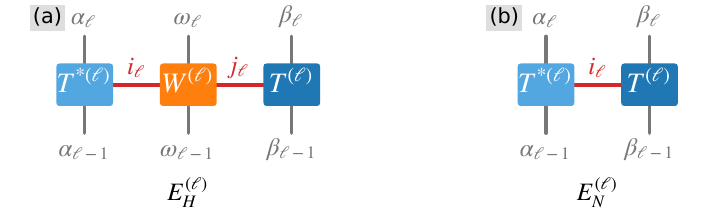}
	\caption{ 
		Diagrammatic representation of the Hamiltonian tranfert matrix (a) and norm transfert matrix(b).
	}
	\label{fig:tranfer_matrix}
\end{figure}

The local transfer matrix at site \(\ell\) associated with the Hamiltonian in Eq.~\eqref{eq:ring_mpo} is obtained by inserting the local MPO tensor \(W^{(\ell)}\) between \(T^{(\ell)*}\) and \(T^{(\ell)}\) (see Fig.~\ref{fig:tranfer_matrix}(a)):
\begin{align}
&\left[E_{\mathcal{H}}^{(\ell)}\right]_{\alpha_{\ell-1}\omega_{\ell-1}\beta_{\ell-1},
	\alpha_\ell\omega_\ell\beta_\ell}
\nonumber\\
&=
\sum_{i_\ell,j_\ell}
\left[T^{(\ell)}_{i_\ell}\right]^*_{\alpha_{\ell-1},\alpha_\ell}
\left[W^{(\ell)}_{i_\ell,j_\ell}\right]_{\omega_{\ell-1},\omega_\ell}
\left[T^{(\ell)}_{j_\ell}\right]_{\beta_{\ell-1},\beta_\ell}.
\label{eq:transfer_H}
\end{align}
Juxtaposed indices such as \(\alpha_\ell\omega_\ell\beta_\ell\) are understood as a single composite index. Accordingly, \(E_{\mathcal{H}}^{(\ell)}\) is a matrix of dimension \(\chi_{\ell-1}^2D\times\chi_\ell^2D\), acting on the composite bra--MPO--ket virtual space.

For the identity operator, for which the local physical action reduces to \(\delta_{i_\ell,j_\ell}\), this construction gives the local transfer matrix associated with the norm (see Fig.~\ref{fig:tranfer_matrix}(b)):
\begin{equation}
\left[E_{\mathcal{N}}^{(\ell)}\right]_{\alpha_{\ell-1}\beta_{\ell-1},
	\alpha_\ell\beta_\ell}
=
\sum_{i_\ell}
\left[T^{(\ell)}_{i_\ell}\right]^*_{\alpha_{\ell-1},\alpha_\ell}
\left[T^{(\ell)}_{i_\ell}\right]_{\beta_{\ell-1},\beta_\ell}.
\label{eq:transfer_N}
\end{equation}

Transfer matrices can also be combined in cyclic products, with site indices understood modulo \(N\). The transfer matrix from \(m\) to \(n\) is defined as the ordered product of local transfer matrices
\begin{equation}
\mathop{\bigcirc}_{\ell=m \ldots n} E^{(\ell)}
:=
E^{(m)}
E^{(m+1)}
\cdots
E^{(n-1)}
E^{(n)}\,,
\label{eq:cyclic_contraction}
\end{equation}
such that expectation values are expressed as
\begin{equation}
\braket{\phi | H | \phi}
=
\operatorname{Tr}\!\left(
\mathop{\bigcirc}_{\ell=1}^{N}
E_{\mathcal{H}}^{(\ell)}
\right),
\end{equation}
while the norm is given by
\begin{equation}
\braket{\phi | \phi}
=
\operatorname{Tr}\!\left(
\mathop{\bigcirc}_{\ell=1}^{N}
E_{\mathcal{N}}^{(\ell)}
\right).
\end{equation}

\subsection{Generalized eigenvalue problem}
\begin{figure}[!b]
	\centering
	\includegraphics[width=\linewidth]{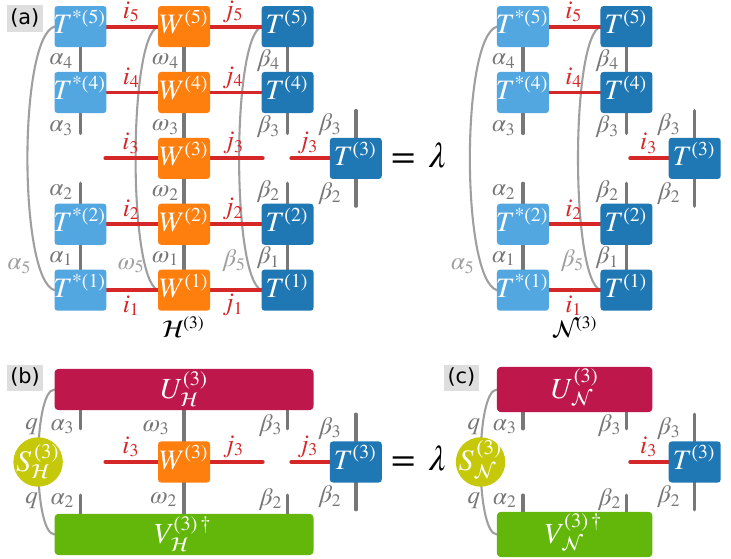}
	\caption{ 
		Diagrammatic representation of the the tensor ring contractions with $N=5$ at site $\ell=3$. (a) generalized eigen value problem Eq.~\eqref{eq:generalized}, (b) compact SVD-based representation of the local Hamiltonian $\mathcal H^{(3)}$ as in Eq.~\eqref{eq:svd_h}, (c) singular value decomposition of the local norm $\mathcal N^{(3)}$ as in Eq.~\eqref{eq:svd_n}.
	}
	\label{fig:generalized_eigh}
\end{figure}

With this notation, the ground state is approximated within the tensor-ring ansatz by minimizing the Rayleigh quotient over the tensor cores:
\begin{equation}
E[\phi]
=
\frac{\braket{\phi|H|\phi}}{\braket{\phi|\phi}},
\qquad
E_0
\simeq
\min_{\{T^{(\ell)}\}} E[\phi].
\end{equation}
Since the wavefunction depends multilinearly on the tensor cores, this defines a nonlinear optimization problem in the full set of variational parameters. DMRG \cite{White92,White93,Schollwock11} addresses this problem by optimizing one core at a time while keeping all others fixed. The optimization proceeds sequentially around the ring and is repeated over successive sweeps until convergence.
At site \(\ell\), the local variational problem reduces to a generalized eigenvalue equation involving the effective norm matrix \(\mathcal{N}^{(\ell)}\) and effective Hamiltonian \(\mathcal{H}^{(\ell)}\). 
The core \(T^{(\ell)}\) is obtained as the corresponding lowest-energy eigenvector (see Fig.~\ref{fig:generalized_eigh}(a)):
\begin{equation}
\mathcal{H}^{(\ell)}_{i_\ell,j_\ell} T^{(\ell)}_{j_\ell}
=
\lambda\,\mathcal{N}^{(\ell)} T^{(\ell)}_{i_\ell}\,.
\label{eq:generalized}
\end{equation}
The effective norm matrix \(\mathcal{N}^{(\ell)}\) is obtained by cyclically contracting all norm transfer matrices except that at site \(\ell\), i.e. from site \(\ell+1\) around the ring to site \(\ell-1\):
\begin{equation}
\mathcal{N}^{(\ell)}
=
\mathop{\bigcirc}_{k\neq \ell}
E_{\mathcal{N}}^{(k)}.
\label{eq:N}
\end{equation}
Similarly, we define the Hamiltonian environment \(\mathcal{H}^{(\ell)}_{\circ}\) by cyclically contracting all Hamiltonian transfer matrices except that at site \(\ell\):
\begin{equation}
\mathcal{H}^{(\ell)}_{\circ}=
\mathop{\bigcirc}_{k\neq \ell}
E_{\mathcal H}^{(k)},\\
\label{eq:W}
\end{equation}
The effective Hamiltonian \(\mathcal{H}^{(\ell)}\) is thus obtained by contracting the environment \(\mathcal{H}^{(\ell)}_{\circ}\) with the local MPO core \(W^{(\ell)}\) over their MPO virtual indices, as illustrated in Fig.~\ref{fig:generalized_eigh}(a) on the left:

\begin{equation}
\mathcal{H}^{(\ell)}_{i_\ell,j_\ell}
=
\mathcal{H}^{(\ell)}_{\circ} W^{(\ell)}_{i_\ell,j_\ell}.
\label{eq:H}
\end{equation}

For tensor trains, the open virtual boundaries allow the state to be brought into a canonical gauge. In this form, the effective norm matrix at the optimization site is the identity, and the local variational problem reduces to an ordinary eigenvalue problem~\cite{Schollwock11}. For tensor rings, the virtual indices form a closed loop, so such a global canonical gauge is generally not available and the effective norm matrix remains nontrivial~\cite{Verstraete04,Pippan10}. The closed geometry also increases the cost of the environment contractions. For fixed physical and MPO dimensions, standard DMRG contractions for tensor trains scale as \(\mathcal{O}(\chi^3)\), whereas the corresponding tensor-ring contractions of Eq.\eqref{eq:N}-\eqref{eq:W} scale as \(\mathcal{O}(\chi^6)\). 

Remarkably, however, long products of transfer matrices exhibit a low-rank structure~\cite{Pippan10}. This allows $\mathcal N^{(\ell)}$ and $\mathcal{H}^{(\ell)}_{\circ}$ to be approximated by a small subset of singular vectors $p_{\mathcal N}^{(\ell)}$ and $p_{\mathcal H}^{(\ell)}$ respectively. We obtain the following decomposition as shown in Fig.~\ref{fig:generalized_eigh}(b),
\begin{align}
\mathcal{N}^{(\ell)}
&=
\sum_{q=1}^{p_{\mathcal N}^{(\ell)}}
S_{\mathcal N,q}^{(\ell)}
U_{\mathcal N,q}^{(\ell)}
\left(V_{\mathcal N,q}^{(\ell)}\right)^\dagger,
\label{eq:svd_n}\\
\mathcal{H}_{\circ}^{(\ell)}
&=
\sum_{q=1}^{p_{\mathcal H}^{(\ell)}}
S_{\mathcal H,q}^{(\ell)}
U_{\mathcal H,q}^{(\ell)}
\left(V_{\mathcal H,q}^{(\ell)}\right)^\dagger.
\label{eq:svd_h}
\end{align}
The singular triplets are computed using a randomized SVD algorithm~\cite{Halko11}, retaining the smallest ranks $p_{\mathcal N}$ and $p_{\mathcal H}$ required to reach a prescribed truncation accuracy $\epsilon_\text{SVD}=10^{-12}$, see Sec.~\ref{sec:rsvd}. The low-rank structure of the environment is a key ingredient of the method introduced in Ref.~\cite{Pippan10}, as it makes the treatment of tensor rings computationally tractable despite their periodic geometry. A more detailed discussion of this low-rank behaviour, including the number of singular components required in practice, is given in Sec.~\ref{sec:low-rank}.

Although the low-rank approximation reduces the size of the environment matrices, recomputing their SVD at every optimization site remains expensive. Constructing an environment requires contracting $\mathcal{O}(N)$ transfer matrices; repeating this procedure independently at each of the $N$ sites therefore leads to an overall cost of $\mathcal{O}(N^2)$ per sweep. To reduce this overhead, Ref.~\cite{Rossini11} introduces a block of $n_{\mathrm{open}}$ consecutive sites,
\begin{equation}
\mathcal{B}_\ell
=
\{\ell,\ell+1,\ldots,\ell+n_{\mathrm{open}}-1\},
\end{equation}
with site indices understood modulo $N$.
Rather than constructing a separate environment for every site, the norm and Hamiltonian environments associated with the block are computed only once. For example, the norm environment admits the low-rank decomposition
\begin{equation}
\mathop{\bigcirc}_{k\notin\mathcal{B}_\ell}
E^{(k)}_\mathcal N
=
\sum_{q=1}^{p_{\mathcal N}^{(\mathcal{B}_\ell)}}
S_{\mathcal N,q}^{(\mathcal{B}_\ell)}
U_{\mathcal N,q}^{(\mathcal{B}_\ell)}
\left(V_{\mathcal N,q}^{(\mathcal{B}_\ell)}\right)^\dagger,
\label{eq:window_environment}
\end{equation}
The same decomposition is reused for all sites in the block. As each site inside the block is optimized, the decomposition is updated for the next site. After the block is completed, a new decomposition as in Eq.~\eqref{eq:window_environment} is computed for the next block.

The environment can therefore be updated locally without recomputing its full low-rank decomposition at every site. Since a sweep contains approximately $N/n_{\mathrm{open}}$ such blocks, the cost of constructing the environments is reduced from $\mathcal{O}(N^2)$ to
$\mathcal{O}\!\left(N^2/n_{\mathrm{open}}\right)$. We observe (see App.~\ref{app:env_scaling}) that choosing \(n_{\mathrm{open}}=2N/3\) increases the required environment rank $p_{\mathcal N}^{(\mathcal{B}_\ell)}$ only moderately, by roughly a factor of two, while reducing the overall environment-construction cost to linear scaling in \(N\).

\subsection{Iterative solver}
\label{sec:solvers}
Solving the generalized eigenvalue problem in Eq.~\eqref{eq:generalized} is carried out, as in DMRG, using a matrix-free iterative solver which never fully constructs the effective operators $\mathcal H^{(\ell)}$ and $\mathcal N ^{(\ell)}$. Rather, the solver only require the action of $\mathcal H ^{(\ell)}$ and $\mathcal N ^{(\ell)}$ on trial vectors. These applications have a reduced cost of $\mathcal{O}\left(p\chi^3\right)$ and avoid the explicit construction and manipulation of matrices of dimension $\chi^2\times\chi^2$.

\begin{figure}
	\centering
	\includegraphics[width=0.49\textwidth]{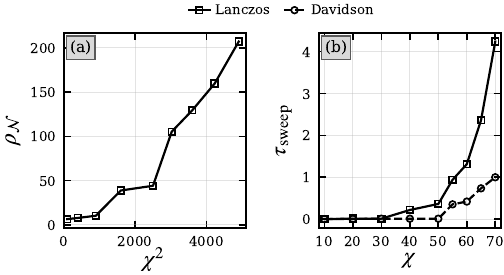}
	\caption{
		Comparison between the Lanczos and Davidson solvers for the local
		generalized eigenvalue problem.
		(a) Number of norm-environment applications $\rho_{\mathcal N}$ for the
		Lanczos solver as a function of $\chi^2$.
		(b) Mean sweep time $\tau_{\mathrm{sweep}}$ as a function of the bond
		dimension $\chi$, normalized by the Davidson sweep time at $\chi=70$.
		The Lanczos solver requires an increasing number of norm-environment
		applications because of the inner linear solves, resulting in a
		substantially larger sweep cost at large bond dimension.
	}
	\label{fig:lanczos_vs_davidson}
\end{figure}

Previous Ring-DMRG implementations~\cite{Pippan10,Rossini11} employed a Lanczos method. For this, the generalized eigenvalue problem from Eq.~\eqref{eq:generalized} is recasted as a standard eigenvalue problem:
\begin{equation}
\mathcal A^{(\ell)} x = \lambda x,
\qquad
\mathcal A^{(\ell)} \equiv \left[\mathcal N^{(\ell)}\right]^{-1}\mathcal H^{(\ell)} .
\label{eq:A}
\end{equation}
Then the usual Krylov-subspace construction is performed on Eq.~\eqref{eq:A}. Starting from an initial vector $v_0$, $\mathcal A^{(\ell)}$  is successively applied to form an $m$-dimensional Krylov space:
\begin{equation}
\mathcal{K}_m
=
\operatorname{span}
\left\{
v_0,
\mathcal{A}^{(\ell)}v_0,
\ldots,
\left[\mathcal{A}^{(\ell)}\right]^{m-1}v_0
\right\}.
\label{eq:A_eigen}
\end{equation}
The eigenvalue problem is then projected and solved onto this reduced space to obtain an approximate eigenpair. Note that each application of $\mathcal A^{(\ell)}$ requires evaluating the action of $\left[\mathcal N^{(\ell)}\right]^{-1}$, which is itself obtained through conjugate-gradient iterations. The method therefore has a nested Krylov structure: the outer Lanczos iteration builds the eigensolver search space, while each application of $\mathcal A^{(\ell)}$ involves an inner conjugate-gradient $\mathcal N^{(\ell)}$. We observe that the mean number of application of $\mathcal N^{(\ell)}$ per conjugate gradient (denoted $\rho_\mathcal{N}$) scales as $\chi^2$ reflecting the worsening of the condition number (see Fig.~\ref{fig:lanczos_vs_davidson}(a)). 
Combined with the $\mathcal{O}(p_\mathcal N\chi^3)$ cost of an individual norm application, this leads to an effective scaling of approximately
$\mathcal{O}(p_\mathcal N \chi^5)$ for the local eigenvalue problem.

To avoid manipulations of $\left[\mathcal{N}^{(\ell)}\right]^{-1}$, we instead employ a tensor-native Davidson method \cite{Ovtchinnikov03,Dorando07} that acts directly on the generalized eigenvalue problem. The method constructs iteratively a search space
\begin{equation}
\mathcal{Q}_m = \operatorname{span}\left \{q_1,\ldots,q_m\right\},
\end{equation}
whose basis vectors are chosen orthonormal with respect to the metric defined by $\mathcal{N}^{(\ell)}$,
\begin{equation}
\left\langle q_a,\mathcal{N}^{(\ell)}q_b\right\rangle = \delta_{ab}.
\label{eq:N_orthogonality}
\end{equation}
Within this subspace, the generalized eigenvalue problem is projected as
\begin{equation}
(\mathcal{H}_s)_{ab} = \left\langle q_a,\mathcal{H}^{(\ell)}q_b\right\rangle,
\qquad
(\mathcal{N}_s)_{ab} = \left\langle q_a,\mathcal{N}^{(\ell)}q_b\right\rangle.
\end{equation}
Because the basis is $\mathcal{N}^{(\ell)}$-orthonormal, $\mathcal{N}_s = \mathds{1}$, and the reduced problem becomes
\begin{equation}
\mathcal{H}_s y = \lambda y.
\end{equation}
After solving this small eigenvalue problem, we denote by $x$ the resulting approximate eigenvector in the original full variational space. We define the residual as
\begin{equation}
r \equiv \mathcal{H}^{(\ell)}x - \lambda\mathcal{N}^{(\ell)}x.
\label{eq:davidson_residual}
\end{equation}
The residual therefore measures the component of the approximate eigenvector that is not captured by the current search space. If its norm is larger than the prescribed convergence threshold, it is orthonormalized and added in the current each space. In this way, the search space is enlarged adaptively in the direction indicated by the current residual until convergence is reached.

In this Davidson solver; each expansion of the search space requires only one direct application of $\mathcal{H}^{(\ell)}$ and $\mathcal{N}^{(\ell)}$ to the new basis vector. The Davidson solver therefore provides a substantial speedup over the Lanczos approach at large $\chi$, as shown in Fig.~\ref{fig:lanczos_vs_davidson}(b), reaching a factor of approximately $5$ at $\chi=70$.

\subsection{Gauging and change of frame}
\label{sec:frame}
A gauge transformation $G$ is an invertible transformation on the cores of the tensor ring that leaves the physical state $\ket{\phi}$ unchanged. For two neighboring tensors, we write
\begin{equation}
\widetilde T_{i_\ell}^{(\ell)} \equiv T_{i_\ell}^{(\ell)} G,
\qquad
\widetilde T_{i_{\ell+1}}^{(\ell+1)} \equiv G^{-1} T_{i_{\ell+1}}^{(\ell+1)},
\end{equation}
so that the contraction across the bond remains invariant,
\begin{equation}
\widetilde T_{i_\ell}^{(\ell)}\widetilde T_{i_{k+1}}^{(\ell+1)} =
T_{i_\ell}^{(\ell)}GG^{-1}T_{i_{\ell+1}}^{(\ell+1)}
=
T_{i_\ell}^{(\ell)}T_{i_{\ell+1}}^{(\ell+1)}.
\end{equation}
Different gauge choices on the virtual bonds of $T^{(\ell)}$ can nevertheless lead to very different effective norm matrices $\mathcal N^{(\ell)}$. A poorly conditioned $\mathcal N^{(\ell)}$ degrades the Davidson solver performances increasing the number of iterations required for convergence.

\begin{figure}
	\centering
	\includegraphics[width=0.4\textwidth]{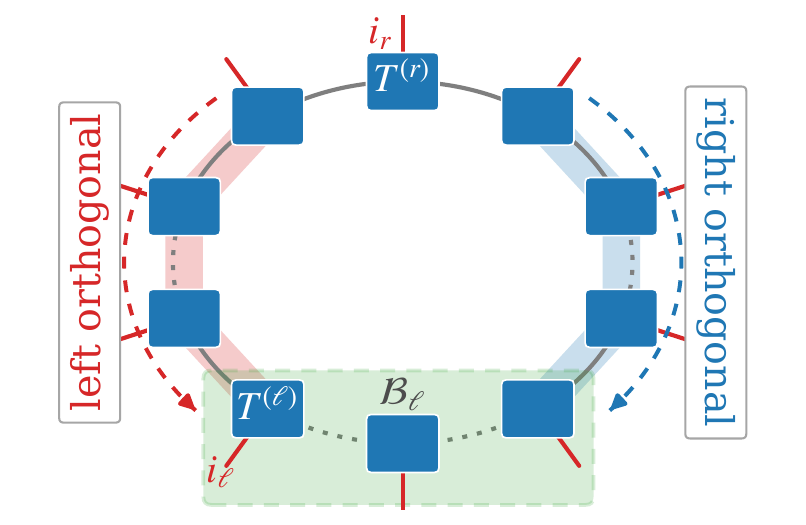}
	\caption{
		Gauging of a tensor ring by left and right orthogonalization.
	}
	\label{fig:gauging}
\end{figure}

Although a global canonical form is unavailable for a tensor ring, a well-conditioned local gauge can still be constructed around the active window, as shown in Fig.~\ref{fig:gauging}. We choose a reference site opposite to the window $\mathcal B_\ell$, denoted by $r=(\ell+N/2)$ (modulo $N$) and perform QR decompositions along the two complementary branches of the ring toward site $\ell$. At each site, the local tensor is replaced by its orthogonal factor $Q$, while the triangular factor $R$ is absorbed into the next tensor along the sweep. This brings the two branches into left- and right-orthogonal forms, respectively. By fixing this gauge, we assure that local optimizations explore different physical states rather than different gauge representations of the same physical state, increasing the stability. During the DMRG sweep through the window (from left to right), the optimized tensors are brought into this form as they are updated: tensors already passed by the sweep are left-orthogonal, while those yet to be optimized remain right-orthogonal.
 The resulting gauge provides a well-conditioned starting point for the environment-whitening procedure introduced below.

\begin{figure}
	\centering
	\includegraphics[width=0.49\textwidth]{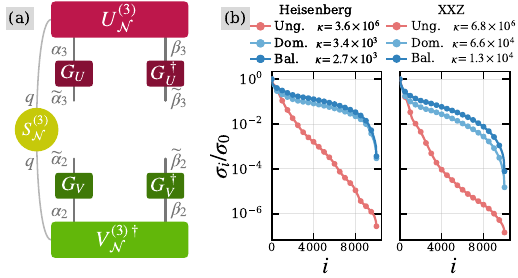}
	\caption{
		Gauging of the norm environment and resulting singular-value spectra.
		(a) Diagrammatic representation of the change of frame applied to the
		compressed norm environment. 
		(b) Normalized singular-value spectrum
		$\sigma_i/\sigma_0$ of the norm environment for the XXZ and Heisenberg
		models in the ungauged (Ung.), dominant (Dom.), and balanced (Bal.)
		frames. The corresponding condition numbers
		$\kappa=\sigma_0/\sigma_{\min}$ are reported in the legend. 
	}
	\label{fig:spectrum_frame}
\end{figure}

To reduce the number of Davidson iterations, we temporarily transform the local problem to a better-conditioned basis during the optimization of core $\ell$. Following Ref.~\cite{Rossini11}, this local change of frame approximately whitens the effective norm matrix $\mathcal N^{(\ell)}$ i.e. brings it closer to the identity and thereby improves its conditioning. The transformation is reverted after the local solve. We first consider the \emph{dominant} frame introduced in Ref.~\cite{Rossini11}. Take the $q=1$ dominant singular triplet of Eq.~\ref{eq:svd_n}, we define the whitening gauge
transformations by
\begin{equation}
G_U^{(\ell)}
=
\left[U_{\mathcal N,1}^{(\ell)}\right]^{-1/2},
\qquad
G_V^{(\ell)}
=
\left[\left(V_{\mathcal N,1}^{(\ell)}\right)^\dagger \right]^{-1/2}.
\label{eq:dominant_whitening}
\end{equation}
These transformations define a new local basis for the tensor $T^{(\ell)}$, acting with $G_U^{(\ell)}$ and $G_V^{(\ell)}$ on its left and right virtual indices, respectively. In this frame, the effective norm matrix becomes

\begin{equation}
\widetilde{\mathcal{N}}^{(\ell)}
=
\left(G_U^{(\ell)}\otimes G_V^{(\ell)}\right)\,
\mathcal{N}^{(\ell)}
\,
\left(\left[G_U^{(\ell)}\right]^\dagger\otimes \left[G_V^{(\ell)}\right]^\dagger \right).
\label{eq:transformed_one_site_metric}
\end{equation}
in Fig.~\ref{fig:spectrum_frame}(a)The same transformation is applied to the Hamiltonian.
Explicitly, the new effective norm components therefore take the form:

\begin{align}
\widetilde U_{\mathcal N,p}^{(\ell)}
&=
G_U^{(\ell)}
U_{\mathcal N,p}^{(\ell)}
\left[G_U^{(\ell)}\right]^\dagger,
\\
\widetilde V_{\mathcal N,p}^{(\ell)}
&=
G_V^{(\ell)}
V_{\mathcal N,p}^{(\ell)}
\left[G_V^{(\ell)}\right]^\dagger.
\end{align}
as shown in Fig.~\ref{fig:spectrum_frame}a.
By construction, $\widetilde U_{\mathcal N,1}^{(\ell)}=\mathds{1}$ and $\widetilde V_{\mathcal N,1}^{(\ell)}=\mathds{1}$. The dominant contribution to $\widetilde{\mathcal N}^{(\ell)}$ is therefore proportional to the identity, while deviations from the identity arise only from the subleading singular components. This generally improves the conditioning of the effective norm matrix. The construction is closely related to the canonical, or Vidal, gauge used for tensor network states~\cite{Vidal04,Tindall23}. 

Note that the dominant-frame approximation is most effective when $S_{\mathcal N,1}^{(\ell)}$ is much larger than the remaining singular values, such that $\mathcal N^{(\ell)}$ is well approximated by a single separable component. If several singular components carry comparable weight, whitening only the dominant component becomes less effective.
To account for all singular components, we introduce a new transformation, which we call the \emph{balanced} frame. We first construct reduced norm environments on the left and right virtual boundaries of Eq.~\eqref{eq:svd_n} by tracing out the opposite boundary,
\begin{align}
L^{(\ell)}
&=
\sum_{q=1}^{p_{\mathcal N}}
S_{\mathcal N,q}^{(\ell)}
\operatorname{Tr}\left(
V_{\mathcal N,q}^{(\ell)}
\right)
U_{\mathcal N,q}^{(\ell)},
\label{eq:left_marginal}\\
R^{(\ell)}
&=
\sum_{q=1}^{p_{\mathcal N}}
S_{\mathcal N,q}^{(\ell)}
\operatorname{Tr}\left(
U_{\mathcal N,q}^{(\ell)}
\right)
V_{\mathcal N,q}^{(\ell)}.
\label{eq:right_marginal}
\end{align}
These matrices define effective norm environments for the left and right virtual boundaries while retaining contributions from all singular components. We use them to construct the whitening transformations
\begin{equation}
G_U^{(\ell)} = \left[L^{(\ell)}\right]^{-1/2},
\qquad
G_V^{(\ell)} = \left[R^{(\ell)}\right]^{-1/2}.
\end{equation}
These transformations are applied to $\mathcal N^{(\ell)}$ as in Eq.~\eqref{eq:transformed_one_site_metric}, yielding the transformed metric $\widetilde{\mathcal N}^{(\ell)}$. The reduced environments $\widetilde L^{(\ell)}$ and $\widetilde R^{(\ell)}$ are then recomputed from $\widetilde{\mathcal N}^{(\ell)}$, and the procedure is iterated until the new norm is sufficiently whitened. In practice, convergence is typically reached in fewer than ten iterations. This iterative balancing provides a more uniform whitening of the full norm environment than the dominant frame.

Figure~\ref{fig:spectrum_frame}(b) compares the conditioning of the norm environment in the different frames for the Heisenberg model and the XXZ model at $\Delta=0.8$~\cite{Yang66}. In both cases, the dominant and balanced frames reduce the condition number by roughly three orders of magnitude compared with the ungauged representation. For the Heisenberg model, both frames perform similarly: the balanced frame only improves the conditioning by an additional factor of about $1.5$. In the XXZ case, several singular components carry significant weight, so whitening only the leading component is less effective. The balanced frame then reduces the condition number by an additional factor of about $5$ compared with the dominant frame. This improved conditioning translates into a reduction of the number of Davidson iterations by a factor $2$.

\section{Two-site truncation}
\label{sec:two_site}

One novelty of this work is the introduction of a two-site DMRG update scheme, which allows the bond dimension to grow dynamically during the optimization. This is particularly useful in the presence of symmetries (here we implement $U(1)$ symmetry, see Appendix~\ref{app:sym}), since it enables the symmetry sectors to expand and avoids trapping the optimization in an overly restricted sector structure. Beyond this, the two-site update provides an efficient route to large bond dimensions. Rather than converging directly at large $\chi$, where each sweep is costly~\cite{Rossini11}, one can first converge at moderate $\chi$ and then progressively increase the bond dimension, reducing the number of sweeps performed at the largest $\chi$.

\begin{figure}
	\centering
	\includegraphics[width=0.48\textwidth]
	{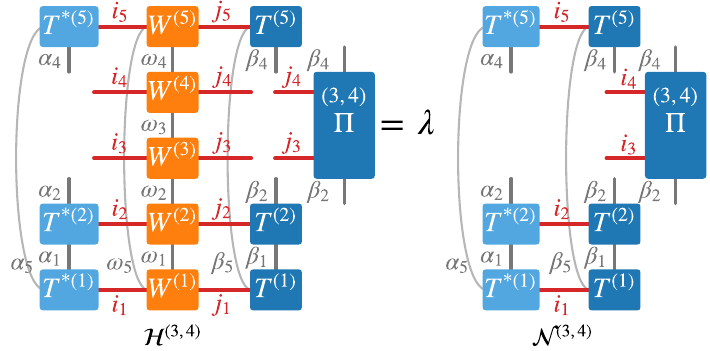}
	\caption{
		Graphical representation of the two-site generalized eigenvalue problem
		\eqref{eq:generalized_2sites}, illustrated for $\ell=3$ and $N=5$.
	}
	\label{fig:eig_two_sites}
\end{figure}

For a two-site update on neighboring sites $\ell$ and $\ell+1$, we define the effective Hamiltonian and norm matrices, $\mathcal H^{(\ell,\ell+1)}$ and $\mathcal N^{(\ell,\ell+1)}$, by cyclically contracting all transfer matrices outside the active two-site block (see Fig.~\ref{fig:eig_two_sites}). The local optimization then reduces to the generalized eigenvalue problem
\begin{equation}
\mathcal{H}^{(\ell,\ell+1)}_{i_\ell,i_{\ell+1},j_\ell,j_{\ell+1}}
\Pi^{(\ell,\ell+1)}_{j_\ell,j_{\ell+1}}
=
\lambda\,
\mathcal{N}^{(\ell,\ell+1)}
\Pi^{(\ell,\ell+1)}_{i_\ell,i_{\ell+1}},
\label{eq:generalized_2sites}
\end{equation}
where for fixed physical indices $i_\ell$ and $i_{\ell+1}$, we denote by $\Pi^{(\ell,\ell+1)}_{i_\ell,i_{\ell+1}}$ the matrix with entries
$\left[\Pi^{(\ell,\ell+1)}_{i_\ell,i_{\ell+1}}\right]_{\alpha_{\ell-1},\alpha_{\ell+1}}$.
The same low-rank environment compression, gauge transformations, and iterative eigensolver introduced for the one-site problem can be applied to Eq.~\eqref{eq:generalized_2sites}. The main additional difficulty for tensor rings lies in the two-site truncation step, in which $\Pi^{(\ell,\ell+1)}$ is split into two tensors, $M^{(\ell)}$ and $M^{(\ell+1)}$, such that
\begin{align}
\Pi^{(\ell,\ell+1)}_{i_\ell,i_{\ell+1}}
=
M^{(\ell)}_{i_\ell}
M^{(\ell+1)}_{i_{\ell+1}}.
\label{eq:two_site_tensor}
\end{align}
If no truncation is performed, the bond dimension between the two sites can increase to as large as $d\chi$. To control this growth, $\Pi^{(\ell,\ell+1)}$ is approximated by two tensors connected by a reduced bond of dimension $\chi$ using an environment-aware truncation.

\subsection{Weighted low-rank approximation problem}

Let $\phi$ denote the state containing the optimized two-site tensor $\Pi^{(\ell,\ell+1)}$, and let $\psi$ denote the state obtained after replacing it by two tensors $M^{(\ell)}$ and $M^{(\ell+1)}$ connected by a bond of dimension $\chi$:
\begin{align}
\phi(i_1,\ldots,i_N)
&=
\operatorname{Tr}\!\left[
T^{(1)}_{i_1}
\cdots
\Pi^{(\ell,\ell+1)}_{i_\ell,i_{\ell+1}}
\cdots
T^{(N)}_{i_N}
\right],
\\
\psi(i_1,\ldots,i_N)
&=
\operatorname{Tr}\!\left[
T^{(1)}_{i_1}
\cdots
M^{(\ell)}_{i_\ell}
M^{(\ell+1)}_{i_{\ell+1}}
\cdots
T^{(N)}_{i_N}
\right].
\end{align}
The optimal truncation minimizes the error in the norm of the full many-body state,
\begin{equation}
\left(
M^{(\ell)},
M^{(\ell+1)}
\right)
=
\underset{
	\operatorname{rank}=\chi
}{\operatorname{arg\,min}}
\,
\left\|
\phi-\psi
\right\|^2.
\label{eq:min_norm}
\end{equation}
The two-site norm environment $\mathcal N^{(\ell,\ell+1)}$ therefore defines a nontrivial metric on the local two-site tensor space. The truncation amounts to minimizing
\begin{equation}
\left(\!M^{(\ell)}_{i_\ell},M^{(\ell+1)}_{i_{\ell+1}}\!\right)
\!=\!
\mathop{\mathrm{arg\,min}}_{\mathrm{rank}=\chi}
\left\|
\Pi^{(\ell,\ell+1)}_{i_\ell,i_{\ell+1}}\!
-\!
M^{(\ell)}_{i_\ell}M^{(\ell+1)}_{i_{\ell+1}}
\right\|_{\!\mathcal N^{(\ell,\ell+1)}}^2 \!.
\label{eq:weighted_low_rank}
\end{equation}
For tensor trains in the canonical gauge, the metric reduces to the identity. By the Eckart--Young theorem~\cite{Eckart36}, the optimal truncation is then obtained from the truncated singular value decomposition of $\Pi^{(\ell,\ell+1)}$. For tensor rings, however, the metric couples the left and right virtual degrees of freedom, so the optimal truncation is no longer obtained from a single SVD and instead takes the form of a \emph{weighted low-rank approximation}. This optimization problem is nonconvex and can exhibit nonglobal local minima~\cite{Srebro03,Markovsky12}; finding the global optimum is NP-hard~\cite{Gillis2011}.

\subsection{Truncation with a separable metric}

When the two-site norm metric is sufficiently separable~\cite{Evenbly18,Vanhecke21,Sinha22}, the change of frame introduced above can approximately whiten $\mathcal N^{(\ell,\ell+1)}$. In this case, the truncation can be approximately performed using an ordinary SVD in the transformed frame. Let $G_U^{(\ell,\ell+1)}$ and $G_V^{(\ell,\ell+1)}$ denote the corresponding whitening transformations acting on the left and right outer virtual bonds. The two-site tensor is transformed as
\begin{equation}
\widetilde{\Pi}^{(\ell,\ell+1)}_{i_\ell,i_{\ell+1}}
=
\left[
\left(G_U^{(\ell,\ell+1)}\right)^\dagger
\right]^{-1}
\Pi^{(\ell,\ell+1)}_{i_\ell,i_{\ell+1}}
\left[
\left(G_V^{(\ell,\ell+1)}\right)^\dagger
\right]^{-1}.
\label{eq:Pi_whitened}
\end{equation}
Its rank-$\chi$ approximation is then obtained from the truncated SVD of $\widetilde{\Pi}^{(\ell,\ell+1)}$. After truncation, the two tensors are transformed back to the original frame,
\begin{equation}
M^{(\ell)}_{i_\ell}
=
\left(G_U^{(\ell,\ell+1)}\right)^\dagger
\widetilde M^{(\ell)}_{i_\ell},
\ \
M^{(\ell+1)}_{i_{\ell+1}}
=
\widetilde M^{(\ell+1)}_{i_{\ell+1}}
\left(G_V^{(\ell,\ell+1)}\right)^\dagger \!.
\label{eq:two_site_backtransform}
\end{equation}

\subsection{Full-metric truncation by alternating least squares}

To further improve the tensors obtained from the truncated SVD in the gauged frame, we introduce an alternating least-squares (ALS) procedure, closely related to the full-environment truncation introduced for tensor networks with closed loops in Ref.~\cite{Evenbly18}. In this approach, the tensors $M^{(\ell)}$ and $M^{(\ell+1)}$ are optimized alternately, with one tensor held fixed while the other is updated. Each update reduces to a linear least-squares problem.

\begin{figure}
	\centering
	\includegraphics[width=0.48\textwidth]{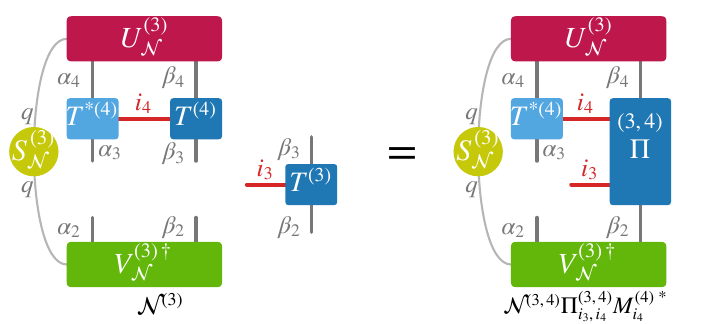}
	\caption{
		Graphical representation of the normal equation
		\eqref{eq:normal_ed} for $\ell=3$, evaluated using the low-rank
		representation of the two-site norm environment $\mathcal N^{(3,4)}$.
	}
	\label{fig:normal_eq}
\end{figure}

In Fig.~\ref{fig:normal_eq}, we illustrate this procedure for the optimization of $M^{(\ell)}$. Starting from the same two-site norm environment $\mathcal N^{(\ell,\ell+1)}$ used in the DMRG step, we construct the corresponding one-site norm matrix $\mathcal N^{(\ell)}$ by absorbing $M^{(\ell+1)}$ and its complex conjugate. Equation~\eqref{eq:weighted_low_rank} then leads to the normal equation:

\begin{equation}
\mathcal N^{(\ell)}
M^{(\ell)}_{i_\ell}
=
\left(M^{(\ell+1)}_{i_{\ell+1}}\right)^*
\mathcal N^{(\ell,\ell+1)}
\Pi^{(\ell,\ell+1)}_{i_\ell,i_{\ell+1}}.
\label{eq:normal_ed}
\end{equation}
An analogous equation is obtained for $M^{(\ell+1)}$ by fixing $M^{(\ell)}$, and the two updates are alternated until the weighted truncation error converges. In practice, convergence is typically reached within fewer than ten iterations. Unlike an SVD for MPS, however, the ALS procedure does not determine the optimal bond dimension $\chi$, and it may be necessary to repeat the ALS for different values of $\chi$ until a prescribed truncation tolerance is reached.

The main practical challenge is to solve the least-squares problem in Eq.~\eqref{eq:normal_ed} using only matrix--vector products. As discussed in Sec.~\ref{sec:solvers}, conjugate-gradient methods become unreliable when the norm metric is poorly conditioned. We therefore use a Galerkin subspace method with a deflated orthogonalization strategy analogous to that employed in our Davidson solver~\cite{Saad03}.

We construct an orthonormal basis $\{q_1,\ldots,q_m\}$ and project the linear system onto the corresponding subspace. The resulting reduced system is solved using a spectral cutoff that removes near-null directions of the metric. The residual is then orthogonalized against the existing basis and added as a new search direction. This procedure is repeated until convergence. The entire solve is performed in the gauged frame, where the norm metric is better conditioned.

\subsection{Numerical experiment}
\label{sec:two_site_dmrg}

\begin{figure}
	\centering
	\includegraphics[width=\columnwidth]{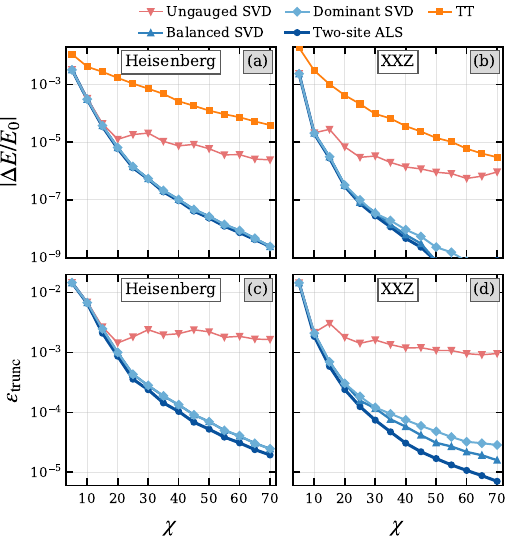}
	\caption{
		Convergence of the two-site Ring-DMRG algorithm for the Heisenberg and XXZ
		models with $N=50$. Panels (a) and (b) show the relative ground-state energy
		error as a function of the bond dimension $\chi$ for the Heisenberg and XXZ
		models, respectively, comparing two-site ALS and the
		ungauged, dominant-frame, and balanced-frame SVD truncations; the corresponding
		tensor-train (TT) results are shown for reference. The reference energies
		$E_{\mathrm{ref}}$ are obtained from TT calculations with bond dimension
		$\chi_{\mathrm{TT}}=3000$. Panels (c) and (d) show the two-site truncation
		error $\epsilon_{\mathrm{trunc}}$ for the same models and truncation schemes.
		The full-metric ALS truncation gives the smallest truncation error,
		with the advantage becoming more pronounced for the XXZ model.
	}
	\label{fig:comp_solver}
\end{figure}

We assess the robustness of the two-site Ring-DMRG algorithm using the same Heisenberg and XXZ models considered in the previous section, with $N=50$. Starting from a N\'eel state at bond dimension $\chi=1$, we progressively increase $\chi$ through two-site updates until the target bond dimension is reached, after which additional sweeps are performed at fixed $\chi$. For comparison, we consider the ungauged SVD together with the dominant-frame, balanced-frame, and ALS truncation schemes.

Panels (a) and (b) of Fig.~\ref{fig:comp_solver} show the relative energy error for the Heisenberg and XXZ models, respectively. The ungauged SVD leads to substantially poorer convergence, whereas the ALS, dominant-frame, and balanced-frame truncations give nearly identical results for the Heisenberg model. For the XXZ model, the dominant-frame truncation is slightly less accurate than the balanced-frame and ALS schemes.

Panels (c) and (d) show the corresponding two-site truncation errors. The difference between the truncation schemes is particularly clear for the XXZ model, where a larger number of environment modes contribute significantly. In this case, ALS gives the smallest truncation error, followed by the balanced-frame and then the dominant-frame approximation. These differences are less pronounced in the energy errors of the first row, since close to a variational optimum the energy error is second order in the error of the state~\cite{Wouters14}. Another application of the two-site scheme, to the addition and compression of tensor rings, is presented in App.~\ref{App:ring_addition}.

\subsection{Time evolution}
\label{sec:time_evol}

We benchmark the two-site scheme on the real-time dynamics of the periodic spin-$1/2$ Heisenberg chain. Starting from a Ring-DMRG ground state $\ket{\phi_0}$, we create a local excitation at site $N/2+1$,
\begin{equation}
\ket{\psi(0)}
=
S_{N/2+1}^z\ket{\phi_0}.
\end{equation}
We then evolve the state under the same periodic Heisenberg Hamiltonian using a second-order Trotter decomposition~\cite{Vidal04,Schollwock11}. The local two-site Hamiltonian term acting on sites $j$ and $j+1$ is
\begin{equation}
h_{j,j+1}
=
J\left(
S_j^z S_{j+1}^z
+
\tfrac{1}{2}S_j^+ S_{j+1}^-
+
\tfrac{1}{2}S_j^- S_{j+1}^+
\right).
\end{equation}
At each time step, the corresponding two-site gates are applied first in forward order,
$(1,2),(2,3),\ldots,(N-1,N),(N,1)$, and then in reverse order,
$(N,1),(N-1,N),\ldots,(2,3),(1,2)$:
\begin{equation}
\begin{split}
e^{-iH\tau}
&\approx
e^{-i\tau h_{1,2}/2}
\cdots
e^{-i\tau h_{N,1}/2}
\\
&\qquad\times
e^{-i\tau h_{N,1}/2}
\cdots
e^{-i\tau h_{1,2}/2}.
\end{split}
\label{eq:tdmrg_step}
\end{equation}
After applying each gate, the resulting two-site tensor $\Pi^{(j,j+1)}$ must be truncated back to the target bond dimension $\chi$. We therefore use the metric-aware truncation schemes introduced above.

\begin{figure}[b]
	\centering
	\includegraphics[width=\linewidth]{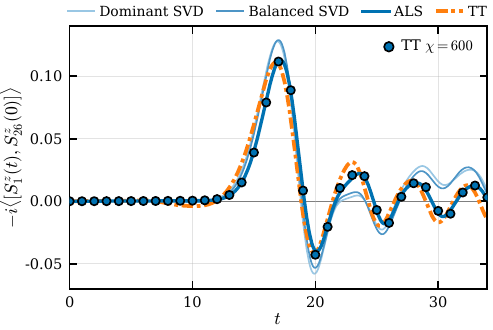}
	\caption{
		Real-time evolution of the retarded spin Green's function
		$G^R_{1,26}(t)=-i\theta(t)\langle [S^z_1(t),S^z_{26}(0)]\rangle$
		for the periodic spin-$1/2$ Heisenberg chain with $N=50$ sites.
		Tensor-ring results at fixed bond dimension $\chi=60$ are shown for the
		dominant-frame, balanced-frame, and ALS truncation schemes. They are compared with
		tensor-train (TT) calculations at $\chi=60$ (orange)
		and $\chi=600$ (blue circles).
	}
	\label{fig:tdmrg_N50}
\end{figure}

In Fig.~\ref{fig:tdmrg_N50}, we consider $N=50$ and $\chi=60$ and compute the retarded spin Green's function between sites $1$ and $26$,
\begin{equation}
G^R_{1,26}(t)
=
-i\theta(t)
\langle
[S^z_1(t),S^z_{26}(0)]
\rangle.
\end{equation}
We compare the dominant-frame, balanced-frame, and full-metric ALS truncations with a tensor-train calculation at $\chi=60$, using a tensor-train calculation at $\chi=600$ as the reference.

For the tensor-train reference, nearest-neighbor gates are treated by standard TEBD in mixed canonical form and truncated by SVD~\cite{Vidal04,Schollwock11}. The periodic gate $(N,1)$ couples the two ends of the train and therefore turns the state into a tensor ring. After applying it, the state is converted back to tensor-train form using App.~\ref{app:ring_to_train} and compressed to the target bond dimension.

The full-metric ALS gives the most accurate dynamics, outperforming the two gauge-based truncations and the tensor-train results at the bond dimensions considered.

\section{FAVORABLE REGIMES FOR RING-DMRG}

\label{sec:regime}

Tensor rings can represent the same tensor with smaller bond dimensions than tensor trains \cite{Handschuh12,Zhao16,Ye18}. In the most favorable case, their bond dimensions are related by
\begin{equation}
\chi_{\rm t} = \chi_{\rm r}^2,
\end{equation}
where $\chi_{\rm t}$ and $\chi_{\rm r}$ are the bond dimensions of the tensor train and tensor ring, respectively (see App.~\ref{app:ring_to_train}). The advantage of Ring-DMRG depends therefore on two parameters: the ring-to-train cost, and the number of singular components in the environment $p_\mathcal{N}$ and $p_\mathcal{H}$. Standard DMRG has a leading cost
$\mathcal{O}(\chi_{\rm t}^3)$, whereas the Ring-DMRG formulation
developed here scales as
$\mathcal{O}(p\chi_{\rm r}^3)$ (see App.~\ref{app:algorithmic_cost}), where $p$ is the typical value of $p_\mathcal{N}$ and $p_\mathcal{H}$. Ignoring prefactors, Ring-DMRG is favored when
\begin{equation}
p\chi_{\rm r}^3
\lesssim
\chi_{\rm t}^3.
\end{equation}

The two quantities can vary essentially independently. In particular, a
tensor ring may have a low-rank environment while remaining expensive
to represent as a tensor train. For instance, we construct (App.~\ref{app:independent_rank_examples}) an explicit family
for which the exact environment rank can be varied from
$p_{\mathcal N}=2$ up to $p_{\mathcal N}=\chi_{\rm r}^2$, while the
tensor-train bond dimension remains fixed at
$\chi_{\rm t}=\chi_{\rm r}^2$.

Conversely, a tensor ring may have a large environment rank without any
ring-to-train overhead. A simple example is the generalized GHZ state
\cite{Vanderstraeten19},
\begin{equation}
|\mathrm{GHZ}_q\rangle
=
\sum_{a=1}^{q}|a\rangle^{\otimes N},
\end{equation}
which admits both tensor-ring and tensor-train representations with bond
dimension $q$. Thus $\chi_{\rm t}=\chi_{\rm r}=q$, while the norm environment has rank $p_{\mathcal N}=q$, since the transfer matrix contains exactly $q$ linearly independent components, one for each GHZ state $|a\rangle$.
\subsection{Low-rank environment and correlation length}
\label{sec:low-rank}

The emergence of low-rank environments is a consequence of the
correlation length of the ring \cite{Pirvu11,Zauner15}. This is particularly clear when considering a
translationally invariant ring whose norm transfer matrix $E_{\mathcal N}$ is diagonalizable:
\begin{equation}
E_{\mathcal N}
\simeq
\sum_{q=1}^{p_{\mathcal N}}
\lambda_q
U_q
V_q,
\end{equation}
with
$|\lambda_1|>|\lambda_2|>\dots$. The eigenspectrum of $E_{\mathcal N}$ controls the decay of correlators: take a local operator $O$ and two sites separated by a distance $L$. The two-point correlator verifies
\begin{equation}
\langle O_i O_{i+L}\rangle
=
\frac{
	\operatorname{Tr}\!\left(
	E_O \left[E_{\mathcal N}\right]^{L-1}
	E_O \left[E_{\mathcal N}\right]^{N-L-1}
	\right)
}{
	\operatorname{Tr}\!\left(\left[E_{\mathcal N}\right]^{N}\right)
},
\end{equation}
where $E_O$ denotes the transfer matrix with an insertion of $O$. Powers of $E_{\mathcal N}$ exponentiate the eigenvalues so that $\langle O_i O_{i+L}\rangle$ takes the form
\begin{equation}
\langle O_i O_{i+L}\rangle
=
\frac{
	\displaystyle
	\sum_{q,r} c_{q,r}
	\lambda_q^{\,L-1}
	\lambda_r^{\,N-L-1}
}{
	\displaystyle
	\sum_s \lambda_s^{\,N}
}.
\end{equation}
The contribution of each mode $q>1$ is therefore exponentially suppressed over a length scale $\xi_q$:
\begin{equation}
\left|\frac{\lambda_q}{\lambda_1}\right|^L
=
\exp\left(-\frac{L}{\xi_q}\right),
\qquad
\xi_q
\equiv
-\frac{1}{
	\log\left|\lambda_q/\lambda_1\right|
}.
\end{equation}
We refer to the \emph{correlation length} $\xi$ as the largest correlation length $\xi_2$.

For a non-translationally invariant ring there is no single transfer matrix
and therefore no unique spectrum $\{\lambda_q\}$. We instead consider the
environment over $L$ consecutive sites,
\begin{equation}
\mathcal E_\ell(L)
=
\mathop{\bigcirc}_{r=\ell}^{\ell+L-1}E_{\mathcal N}^{(r)},
\label{eq:prod_transfer}
\end{equation}
with singular values
$S_{1}^{(\ell)}(L)\geq S_{2}^{(\ell)}(L)\geq\cdots$. In analogy with the translationally
invariant case, we characterize its typical decay empirically by
\begin{equation}
\frac{S_{2}^{(\ell)}(L)}
{S_{1}^{(\ell)}(L)}
\sim
\exp\left(-\frac{L}{\xi}\right).
\label{eq:decay}
\end{equation}
Such an exponential separation of singular directions is rigorously known
for broad classes of random matrix products through multiplicative ergodic
theory \cite{Furstenberg60,Oseledets68}. Here $\xi$ should instead be
understood as an effective decay length characterizing the compressibility of the inhomogeneous environment (see App.~\ref{app:tranfer_prod}).
If the system size is larger than this correlation length $\xi\ll N$, we expect only a few surviving modes and therefore $p_{\mathcal N}$ small \cite{Pirvu11,Pirvu12}. On the contrary, for $N\sim\xi$ many modes can survive and $p_{\mathcal N}$ is typically much larger.

\subsection{Finite-size ring}
\begin{figure}[b]
	\centering
	\includegraphics[width=0.49\textwidth]{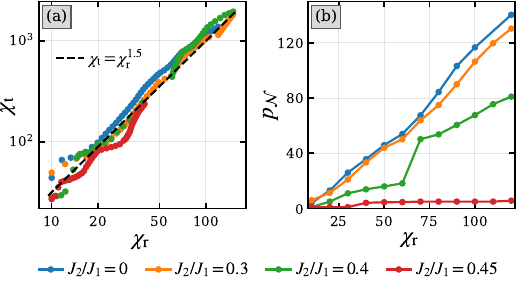}
	\caption{
Comparison of tensor train and tensor ring representations for the
spin-$1/2$ $J_1$--$J_2$ Heisenberg chain.
(a) Tensor train bond dimension $\chi_{\rm t}$ required to reach the same
relative ground-state energy error as a tensor ring of bond dimension
$\chi_{\rm r}$, for $J_2/J_1=0$, $0.30$, $0.40$, and $0.45$.
The dashed line shows the scaling
$\chi_{\rm t}=\chi_{\rm r}^{3/2}$, which is consistent with the numerical
data (see App.~\ref{App:J1J2} for the data).
(b) Mean retained rank $p_{\mathcal N}$ of the norm environment as a function
of the tensor-ring bond dimension $\chi_{\rm r}$ for the same values of
$J_2/J_1$. As $J_2$ increases, the correlation length decreases, causing the environment singular values to decay faster and fewer components to be retained.
	}
	\label{fig:J1J2}
\end{figure}

For a finite system of size \(N=200\), we illustrate the potential gain of Ring-DMRG using the $J_1$-$J_2$ Heisenberg model setting $J_1=1$. The model undergoes a transition from a
Luttinger liquid phase for $J_2<J_c\approx0.241$ to a gapped phase
\cite{Affleck89,Nomura94} and the dimerized Majumdar--Ghosh point  at $J_2=1/2$ \cite{Majumdar69}. We consider four cases $J_2=0$, $0.3$, $0.4$ and $0.45$. In all four cases, we see an approximate scaling $\chi_\mathrm{t}=\chi_\mathrm{r}^{1.5}$ (see Fig.~\ref{fig:J1J2}(a)). However, the environment compressibility depends strongly on $J_2$. As shown in Fig.~\ref{fig:J1J2}(b), $p_{\mathcal N}$ increases rapidly with $\chi$ for $J_2=0$ and $0.3$, exhibits a more moderate increase for $J_2=0.4$, and remains approximately constant at $p_{\mathcal N}=5$ for $J_2=0.45$ over the range of $\chi$ considered. Therefore, Ring-DMRG offers a substantial advantage in the gapped phase but this becomes more limited inside the Luttinger liquid phase.

This observation is consistent with the correlation length extracted from the spin gaps reported in Ref.~\cite{Saha21}. In the gapped phase, $\xi$ can be estimated as:
\begin{equation}
\xi \sim \frac{v}{\Delta_s},
\end{equation}
where \(\Delta_s\) is the spin gap and \(v\) is the spin velocity \cite{Okunishi01}. Ref.~\cite{Saha21} reports $\Delta_s\simeq 7.4\times10^{-4}$ at $J_2/J_1=0.30$ and $\Delta_s\simeq 3.0\times10^{-2}$ at $J_2/J_1=0.40$, while its data at $J_2/J_1=0.45$ give $\Delta_s\simeq 0.11$. Taking $v\sim J_1a$ and setting $J_1=a=1$, these values correspond to correlation lengths of order $10^3$, $30$, and $10$ sites, respectively.

\subsection{Thermodynamic limit of critical periodic systems}

We now turn to the study of critical systems in the thermodynamic limit ($N\to\infty$), where connected correlations decay algebraically with distance \cite{Giamarchi04}. In this setting, we find that Ring-DMRG is particularly advantageous for computing quantities that remain sensitive to the global ring geometry. Examples include long-distance correlation functions \cite{Pirvu11}, spin and charge stiffnesses \cite{Rossini11}, persistent currents in flux-threaded rings \cite{Rizzi14}, many-body polarization and Berry phases defined through flux insertion \cite{Watanabe18}, as well as the Casimir energy, central charge, and scaling dimensions \cite{VanDamme21}.

Approximating the system by a tensor ring introduces two distinct infrared cutoffs. The first is the finite ring circumference $N$, which gives rise to the usual finite-size effects in observables. The second arises from the finite bond dimension $\chi$, which induces an effective finite-entanglement correlation length $\xi_\chi$ \cite{Tagliacozzo08,Pollmann09,Vanhecke19}. This correlation length scales with the bond dimension as
\begin{equation}
\xi_\chi \sim \chi^\kappa,
\end{equation}
where $\kappa$ is determined by the underlying critical theory ($\kappa\approx 1.338$ for the spin-$1/2$ Heisenberg model). For a critical system represented by a tensor ring of size $N$ and bond dimension $\chi$, we observe that the singular-value decay described by Eq.~\eqref{eq:decay} reflects the combined effects of these two infrared cutoffs (see App.~\ref{app:tranfer_prod}):
\begin{equation}
\frac{S_2^{(\ell)}(L)}
{S_1^{(\ell)}(L)}
\sim
\exp\left[
-\underbrace{
	\left(
	\frac{b_1}{N}
	+
	\frac{b_2}{\xi_\chi}
	\right)
}_{1/\xi}L
\right],
\label{eq:decay_fss}
\end{equation}
where $b_1$ and $b_2$ are nonuniversal coefficients that may depend, for example, on the gauge choice. For the norm environment $\mathcal N^{(\ell)}$ introduced in Eq.~\eqref{eq:svd_n}, the transfer product contains $L=N-1$ sites. Neglecting terms of order $1/N$, one gets:
\begin{equation}
\frac{S_{\mathcal N,2}^{(\ell)}}
{S_{\mathcal N,1}^{(\ell)}}
\sim
\exp\left[
-b_2\frac{N}{\xi_\chi}
\right].
\label{eq:decay_rate}
\end{equation}
Approaching the thermodynamic critical state of Eq.~\eqref{eq:decay_rate} requires some care,
since the limits $N\to\infty$ and $\chi\to\infty$ do not commute \cite{Pirvu12}. At fixed $\chi$, the correlation length $\xi_\chi$ remains finite, and
taking $N\to\infty$ exponentially suppresses the subleading transfer
directions: the environments are then
asymptotically dominated by a single transfer mode $p_\mathcal{N} = p_\mathcal{H}=1$. This corresponds to the
\emph{finite-entanglement} regime. Although the global finite-ring state is no longer faithfully represented in this regime, observables probing length scales much smaller than $\xi_\chi$ can still accurately capture the local critical physics \cite{Pirvu12}. This motivates the use of the variational uniform matrix product state (VUMPS) method \cite{Milsted13,Zauner15,ZaunerStauber18}, which works directly in the thermodynamic limit and therefore eliminates finite-size effects associated with the ring circumference.

For observables that remain sensitive to the global periodic geometry,
however, we must take the limit $N\to\infty$ while increasing the bond dimension such that $N<\xi_\chi$. In this case the ring is not affected by the finite bond dimension cutoff, this is the \emph{finite-size} regime. The crossover between the finite-entanglement and finite-size regimes occurs when the two infrared length scales become comparable, $\xi_\chi \sim N$ defining a characteristic bond dimension,
\begin{equation}
\chi_c(N)\sim N^{1/\kappa},
\end{equation}
Scaling $\chi$ together with $N$ such that $\chi \gtrsim \chi_c(N)$ ensures that the finite-entanglement length remains at least of order the ring circumference, allowing correlations to retain sensitivity to the global periodic geometry as the thermodynamic limit is approached.

\begin{figure}
\centering
\includegraphics[width=\linewidth]{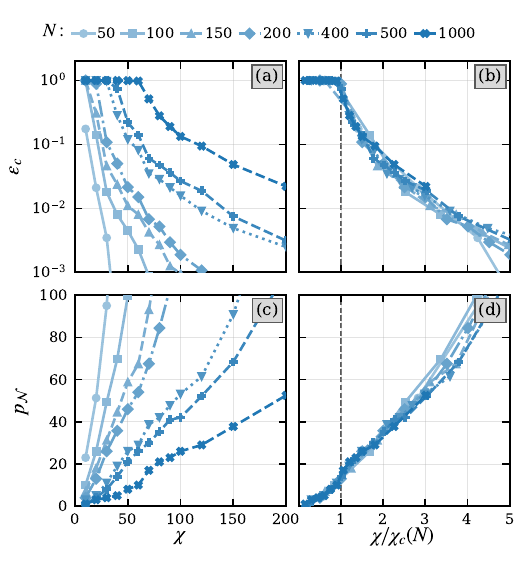}
\caption{
		Finite-size--finite-entanglement collapse of the norm-environment rank and
		of a ring-sensitive correlation function for the periodic spin-$1/2$
		Heisenberg chain. Panel (a) shows the relative error
		$\varepsilon_c=|C_{\rm opp}^{zz}(N)-C_{\rm ref}|/|C_{\rm ref}|$ of the
		opposite-side correlator and panel (c) the mean retained norm-environment rank
		$p_{\mathcal N}$, as a function of $\chi$ and for several system sizes $N$. Note that $C_{\rm ref}$ is obtained by DMRG with $\chi_{t}=3000$. Both $\varepsilon_c$ and $p_\mathcal{N}$ are plotted against the rescaled bond dimension
		$\chi/\chi_c(N)$ in panels (b) and (d). The crossover bond dimension
		is defined from the finite-entanglement correlation length as
		$\chi_c(N)=(N/3.647)^{1/1.338}$, so that
		$\xi_\chi\simeq3.647\,\chi^{1.338}$, and the dashed vertical line marks
		$\chi=\chi_c(N)$, or equivalently $\xi_\chi\simeq N$. After rescaling by
		$\chi_c(N)$ both quantities collapse approximately onto size-independent
		curves.
	}
\label{fig:rank_and_correlator}
\end{figure}

We consider as an example the opposite-site correlator:

\begin{equation}
C_{\rm opp}^{zz}(N)
\equiv
\bigl\langle S_1^z S_{N/2+1}^z \bigr\rangle
-
\bigl\langle S_1^z \bigr\rangle
\bigl\langle S_{N/2+1}^z \bigr\rangle .
\label{eq:opposite_correlator}
\end{equation}
Figure~\ref{fig:rank_and_correlator}(a) shows the relative error in this correlator for several system sizes as a function of the bond dimension. The retained environment rank  $p_\mathcal N$ is given in panel~(c). Rescaling the bond dimension by $\chi_c(N)$ leads to a collapse of the data for different system sizes in panels (b) and (d). This behavior is consistent with the renormalization-group picture of Ref.~\cite{Schneider25}, in which a finite bond dimension introduces an infrared length scale set by $\xi_\chi$. Therefore, achieving a fixed target accuracy in the correlator requires approximately the same ratio $\chi/\chi_c(N)$ and same $p_\mathcal N$ (same $p_\mathcal{H}$) across different system sizes. 

At first sight, it is not obvious that Ring-DMRG retains an advantage in this regime. At fixed $N$, increasing the bond dimension leads to a growing number of relevant environment modes, with $p_{\mathcal N}\sim\chi^2$, as the transfer-matrix spectrum becomes increasingly dense \cite{Vanhecke19}. In App. \ref{app:env_scaling} this quadratic scaling is shown for different system sizes and has also been observed in Ref.~\cite{Rakov16}. If this fixed-$N$ behavior persisted along the thermodynamic-limit trajectory, the Ring-DMRG cost, $\mathcal{O}(p\chi_{\rm r}^3)$, would approach $\mathcal{O}(\chi_{\rm r}^5)$.

\begin{figure}[b]
	\centering
	\includegraphics[width=0.49\textwidth]{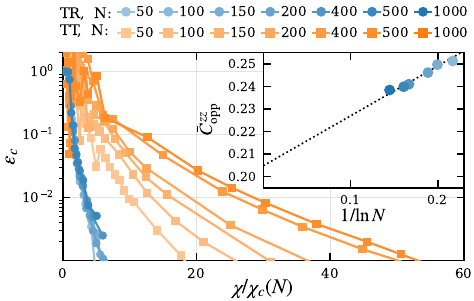}
	\caption{
		Convergence and finite-size scaling of the opposite-side connected spin
		correlator $C_{\rm opp}^{zz}(N)$ defined in
		Eq.~\eqref{eq:opposite_correlator} for the periodic spin-$1/2$ Heisenberg
		chain. Relative correlator error
		$\varepsilon_c$
		for tensor ring (circles) and tensor train (squares) representations are plotted
		as a function of the rescaled bond dimension $\chi/\chi_c(N)$ for several
		system sizes. In the inset, finite-size scaling of the best available tensor-ring
		results is plotted as
		$\bar{C}_{\rm opp}^{zz}(N)=|C_{\rm opp}^{zz}(N)|N/\sqrt{\ln N}$ versus $1/\ln N$. The dotted line shows
		a linear fit, whose extrapolation to $1/\ln N\to0$ corresponds to the
		thermodynamic limit.
	}
	\label{fig:convergence_correlator}
\end{figure}

The collapse above shows, however, that this fixed-$N$ scaling does not control the thermodynamic-limit trajectory at fixed accuracy. To keep a ring-sensitive observable converged as $N$ increases, it is sufficient to scale the bond dimension along a trajectory such that $
\chi_{\rm r}/\chi_c(N)=\mathcal{O}(1).$ Along this trajectory, the retained environment rank $p$ is constant. The resulting Ring-DMRG complexity therefore approaches $\mathcal{O}(\chi_{\rm r}^3)$. By contrast, representing the same state in tensor train form requires a substantially larger bond dimension, leading to a correspondingly higher computational cost. Figure~\ref{fig:convergence_correlator} extends the analysis of
Fig.~\ref{fig:rank_and_correlator} by including the tensor train bond
dimensions required to achieve the same accuracy in the opposite-side
correlator. For $1\lesssim\chi/\chi_c(N)\lesssim5$, the Ring-DMRG data are
well described by an approximately size-independent exponential convergence,
\begin{equation}
\varepsilon_c
\sim
\exp\left[
-a_{\rm r}\frac{\chi}{\chi_c(N)}
\right],
\qquad
a_{\rm r}\simeq1.2.
\end{equation}
By contrast, the tensor-train convergence becomes progressively slower with increasing system size, with an effective rate \(a_{\rm t}(N)\sim N^{-0.4}\). At fixed accuracy, this means that the required tensor-train bond dimension grows faster with \(N\) than the tensor-ring bond dimension. Indeed, since \(\varepsilon_c\sim\exp[-a_{\rm t}(N)\chi_{\rm t}/\chi_c(N)]\), fixed \(\varepsilon_c\) implies \(\chi_{\rm t}\sim\chi_c(N)/a_{\rm t}(N)\sim N^{1/\kappa+0.4}\), whereas \(\chi_{\rm r}\sim\chi_c(N)\sim N^{1/\kappa}\). Eliminating \(N\) therefore gives
\begin{equation}
\chi_{\rm t}\sim\chi_{\rm r}^{\,1+0.4\kappa}
\simeq \chi_{\rm r}^{1.54}
\end{equation}
where \(\kappa\simeq1.338\). This is consistent with our observation that \(\chi_{\rm t}\sim\chi_{\rm r}^{1.5}\) for the Heisenberg model (see Fig.~\ref{fig:N500}).

The inset shows $|C_{\rm opp}^{zz}(N)|N/\sqrt{\ln N}$ as a function of $1/\ln N$. The leading asymptotic behavior of the opposite-site correlator is expected to scale as $|C_{\rm opp}^{zz}(N)|\sim \sqrt{\ln N}/N$ \cite{Barzykin99}, so that this rescaled quantity approaches a finite constant in the thermodynamic limit, with only subleading logarithmic corrections. Extrapolating our Ring-DMRG data to $1/\ln N\to0$ gives a value in close agreement with the theoretical amplitude
$
1/\sqrt{8\pi}\simeq0.20.
$

In App.~\ref{app:algorithmic_cost}, we benchmark the computational performance of our algorithm against that of the ITensor library~\cite{Fishman22}.

\subsection{Central charge from ground-state overlaps}
\label{sec:overlap_central_charge}

\begin{figure}[h!]
	\centering
	\vspace{4pt}
	\includegraphics[width=\columnwidth]{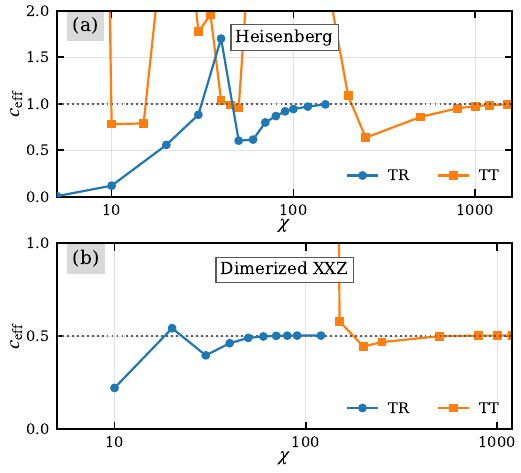}
	\caption{ Central charge extrapolation from ground state overlap.
	The effective central charge $c_{\rm eff}$ is
	extracted from the ground-state overlap between the uniform
	and $q$-Möbius-deformed with $\theta=0.5$ and $q=2$ for the spin-$1/2$ Heisenberg chain (a) and a dimerized XXZ chain (b),
	as a function of bond dimension $\chi$.
	We compare tensor ring (blue) and tensor train (orange)
	representations.}
	\label{fig:ceff}
\end{figure}

As a second example of a quantity that depends on the global structure of the periodic ground state, we consider the recently proposed ground-state-overlap estimator of the central charge \cite{Bai26}. We write the periodic Hamiltonian as
$
\widetilde H=\sum_\ell h_\ell,
$
where $h_\ell$ denotes a local Hamiltonian term assigned to the position $\ell$. The $q$-Möbius deformation is obtained by modulating the local Hamiltonian terms according to
\begin{equation}
\widetilde H_q(\theta) = 
\sum_\ell
\left[
1-\tanh(2\theta)
\cos\left(\frac{2\pi q \ell}{N}\right)
\right]h_\ell,
\end{equation}
where $q$ sets the modulation wave number and $\theta$ controls the deformation strength. Denoting by $|\Omega_0\rangle$ and $|\Omega_q(\theta)\rangle$ the ground states of the uniform and deformed Hamiltonians, respectively, their overlap defines an effective estimate of the central charge \cite{Bai26},
\begin{equation}
c_{\rm eff}
=-\frac{6q}{q^2-1}
\frac{
	\ln\left|
	\langle\Omega_0|\Omega_q(\theta)\rangle
	\right|
}{
	\ln(\cosh\theta)
}.
\label{eq:ceff_overlap}
\end{equation}
Figure~\ref{fig:ceff} compares the convergence of the effective central charge
$c_{\mathrm{eff}}$ with the bond dimension $\chi$ for tensor-ring and
tensor-train representations at $N=400$, with $\theta=0.5$ and $q=2$.
Panel~(a) corresponds to the spin-$1/2$ Heisenberg chain, with an expected
central charge $c=1$, while panel~(b) considers the dimerized spin-$1/2$
XXZ chain with alternating longitudinal couplings $J_a=1.5441$ and
$J_b=1.75$ and uniform transverse coupling $J_\perp=1$, for which
$c=0.5$ is expected at the Ising critical point \cite{Nigam25}.
At this system size, the converged values differ slightly from their
thermodynamic-limit predictions, approaching $c_{\mathrm{eff}}\simeq0.993$
in panel~(a) and $c_{\mathrm{eff}}\simeq0.504$ in panel~(b).

For the Heisenberg chain in panel~(a), the tensor-ring representation
reaches convergence already at $\chi_{\mathrm{r}}\simeq150$, whereas
the tensor-train representation requires approximately
$\chi_{\mathrm{t}}\simeq1500$ to achieve comparable accuracy.
This corresponds to a bond-dimension ratio
$\chi_{\mathrm{t}}/\chi_{\mathrm{r}}\simeq10$.
With a retained Hamiltonian-environment rank
$p_{\mathcal H}\simeq130$, the leading computational costs
$\mathcal{O}(p_{\mathcal H}\chi_{\mathrm{TR}}^3)$ for TR and
$\mathcal{O}(\chi_{\mathrm{t}}^3)$ for TT suggest a potential
speedup of approximately $8$ for the tensor ring representation,
up to implementation-dependent prefactors.
A similar behavior is observed for the dimerized XXZ chain in
panel~(b), where $\chi_{\mathrm{t}}/\chi_{\mathrm{r}}\simeq10$
and $p_{\mathcal H}\simeq80$, yielding an estimated speedup
of approximately $13$.

\section{Randomized SVD and belief propagation}

\label{sec:rsvd}

We briefly describe the randomized singular value decomposition (rSVD) \cite{Halko11} used to construct the low-rank environments and make an analogy with belief propagation. Consider the norm matrix $\mathcal{N}^{(\ell)}$, explicitly forming this matrix requires \(\mathcal O(\chi^4)\) memory, while computing its dense singular value decomposition (SVD) costs \(\mathcal O(\chi^6)\) operations. Instead, rSVD exploits the low-rank structure of \(\mathcal N^{(\ell)}\), reducing the memory requirement to \(\mathcal O(p\chi^2)\) and the computational cost to \(\mathcal O(p\chi^3)\), up to the oversampling and power-iteration factors discussed below.

\begin{figure}[h!]
	\centering
	\includegraphics[width=0.49\textwidth]{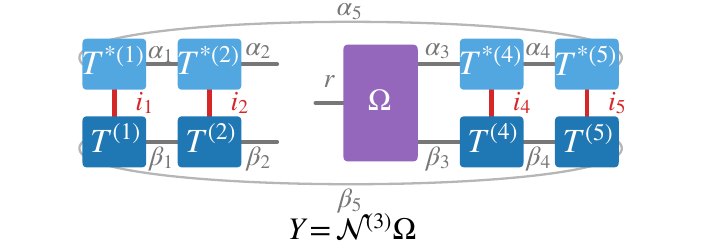}
	\caption{
		Graphical representation of the randomized probing
		Eq.~\eqref{eq:srvd}, illustrated for $\ell=3$ and $N=5$
	}
	\label{fig:rsvd}
\end{figure}

Given a target rank \(p_{\mathcal N}\) for the effective norm, we introduce a small oversampling parameter \(s\) and define \(r=p_{\mathcal N}+s\). We then draw a random Gaussian matrix \(\Omega\) with \(r\) columns and form (see Fig.~\ref{fig:rsvd}):
\begin{equation}
Y=\mathcal N^{(\ell)}\Omega.
\label{eq:srvd}
\end{equation}
The multiplication is carried out directly in tensor-network form, without explicitly constructing \(\mathcal N^{(\ell)}\), at a cost of \(\mathcal O(r d \chi^3)\). The columns of \(Y\) approximately span the dominant left singular subspace of \(\mathcal N^{(\ell)}\). We therefore orthonormalize them, \(Y=QR\), and project \(\mathcal N^{(\ell)}\) onto this reduced subspace:
\begin{equation}
B=Q^\dagger \mathcal N^{(\ell)}.
\end{equation}
Since \(B\) has only \(r\) rows, its SVD is inexpensive compared with that of the full matrix:
\[
B=\widetilde U S V^\dagger.
\]
Retaining the leading \(p_{\mathcal N}\) singular components then yields
\begin{equation}
\mathcal N^{(\ell)}
\simeq
\sum_{q=1}^{p_{\mathcal N}}
S_{\mathcal N,q}^{(\ell)}
U_{\mathcal N,q}^{(\ell)}
\left(V_{\mathcal N,q}^{(\ell)}\right)^\dagger,
\end{equation}
with \(U=Q\widetilde U\).

The accuracy of the randomized approximation can be further improved by performing \(q\) \emph{power iterations}. Starting from \(Y=\mathcal N^{(\ell)}\Omega\), each iteration alternates an application of \(\mathcal N^{(\ell)\dagger}\) in one direction with an application of \(\mathcal N^{(\ell)}\) in the opposite direction. In compact form,
\begin{equation}
Y=
\left(
\mathcal N^{(\ell)}\mathcal N^{(\ell)\dagger}
\right)^q
\mathcal N^{(\ell)}\Omega.
\label{eq:rsvd_power}
\end{equation}
QR orthogonalizations are performed between successive applications to maintain numerical stability. If \(\sigma_j\) is a singular value of \(\mathcal N^{(\ell)}\), its contribution is weighted as \(\sigma_j^{2q+1}\). The power iterations therefore enhance the separation between dominant and subdominant singular directions, improving the quality of the low-rank approximation.

\begin{figure}[h!]
	\centering
	\includegraphics[width=0.49\textwidth]{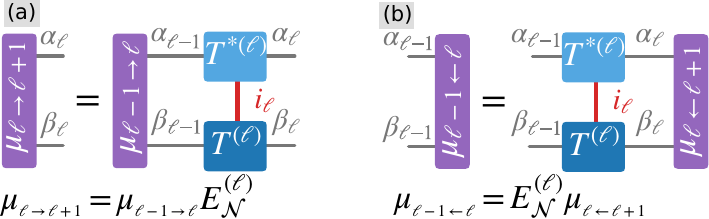}
	\caption{
	 Belief-propagation on a ring. Update of the clockwise (a) and counter-clockwise (b) messages.
	}
	\label{fig:simple_bp}
\end{figure}

This construction is closely related to belief propagation (BP)~\cite{Yedidia05,Alkabetz21,Tindall23}. We recall the BP process by applying it to the doubled norm network \(\langle\phi|\phi\rangle\) of a tensor-ring. The idea of BP is to represent the information propagated through the network by \emph{messages} $\mu$, which are vectors of dimension $\chi^2$ associated with each virtual bond. These messages are iteratively updated by contracting them with the local transfer matrices until they reach a fixed point. We distinguish clockwise messages $\mu_{\ell-1\to\ell}$ and counterclockwise $\mu_{\ell-1\leftarrow\ell}$. These messages are updated around the ring following (see Fig.~\ref{fig:simple_bp}):
\begin{align}
\mu_{\ell\to\ell+1}
&=
\mu_{\ell-1\to\ell}E_{\mathcal N}^{(\ell)}
,
\label{eq:bp_one_forw}
\\
\mu_{\ell-1\leftarrow\ell}
&=
E_{\mathcal N}^{(\ell)}\mu_{\ell\leftarrow\ell+1}.
\label{eq:bp_one_back}
\end{align}
After each update, the messages are normalized.

\begin{figure}[h!!!!]
	\centering
	\includegraphics[width=0.40\textwidth]{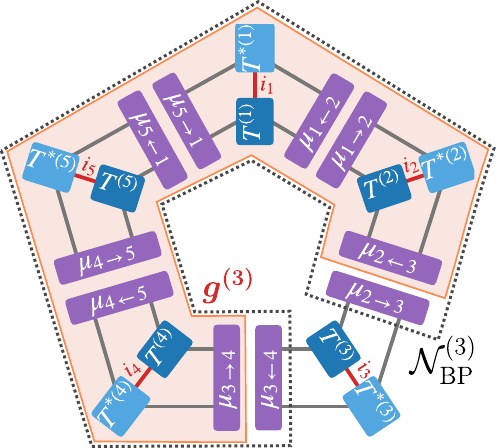}
	\caption{
		Norm of the ring state $\langle\phi|\phi\rangle$ evaluated at the BP fixed point. The projectors defined in Eq.~\eqref{eq:bp_proj} are inserted on every bond. The contractions corresponding to $g^{(3)}$ and $N_{\mathrm{BP}}^{(3)}$ are indicated in red and by dotted lines, respectively.
	}
	\label{fig:bp_norm_simple}
\end{figure}

Repeated sweeps around the ring amount to power iterations of the complete cyclic transfer operator 
\begin{equation}
\mathcal M^{(\ell)}=E_{\mathcal N}^{(\ell+1)}\cdots E_{\mathcal N}^{(\ell)},
\end{equation}
for each $\ell$. After one complete sweep, the clockwise and counterclockwise messages are updated as
\begin{align}
\mu_{\ell\to\ell+1}
&\propto \mu_{\ell\to\ell+1} \mathcal M^{(\ell)},\\
\mu_{\ell\leftarrow\ell+1}
&\propto \mathcal M^{(\ell)} \mu_{\ell\leftarrow\ell+1}.
\end{align}
Upon convergence, the messages approach the dominant left and right eigenvectors of the cyclic transfer operator at each virtual bond. Together, they define a rank-one oblique projector onto the dominant invariant subspace:
\begin{equation}
\begin{gathered}
P^{(\ell)}
=
\frac{
	\mu_{\ell\leftarrow\ell+1}
	\otimes
	\mu_{\ell\to\ell+1}
}{
	\langle \mu_{\ell\to\ell+1}
	| \mu_{\ell\leftarrow\ell+1} \rangle
}
\\[0.5em]
\hspace{2.5em}
\includegraphics[scale=0.7]{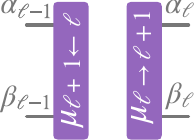}
\end{gathered}
\label{eq:bp_proj}
\end{equation}
For compactness we suppose ${\langle{\mu_{\ell\to\ell+1}}|\mu_{\ell\leftarrow\ell+1}\rangle}=1$. The BP approximation consist of inserting these projectors $P^{(\ell)}$ at every virtual bond. Equivalently, this replaces the original norm transfer-matrix network by a network
of reduced scalar transfer factors,
\begin{equation}
\begin{gathered}
\tilde E_{\mathcal N}^{(\ell)}
=
\mu_{\ell-1\to\ell}
E_{\mathcal N}^{(\ell)}
\mu_{\ell\leftarrow\ell+1}.
\\[0.5em]
\hspace{3em}
\includegraphics[scale=0.7]{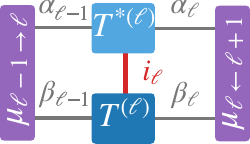}
\end{gathered}
\label{eq:bp_reduced}
\end{equation}
The norm of the ring $\langle\phi|\phi\rangle$ can therefore be approximated (Fig.~\ref{fig:bp_norm_simple}) by:
\begin{equation}
\langle\phi|\phi\rangle \approx \mathrm{Tr}\left(\prod_{k=1}^{N} \tilde E_{\mathcal N}^{(k)}\right).
\end{equation}

\begin{figure}[h!]
	\centering
	\includegraphics[width=0.45\textwidth]{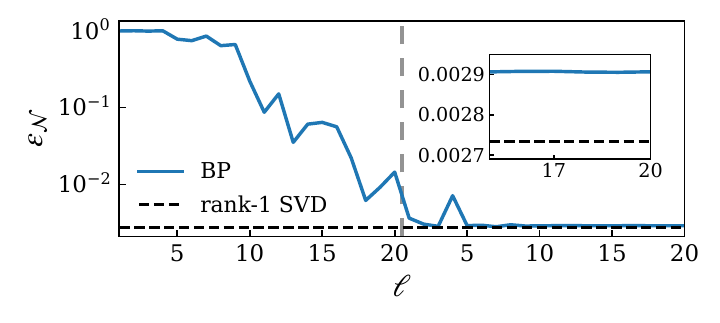}
	\caption{ Convergence of BP on a ring.
For a random tensor ring with $N = 20$ and bond dimension
$\chi = 10$, the relative error $\varepsilon_N$ in the BP approximation
of $\mathcal{N}^{(1)}$ is shown as a function of the site-update index.
Two complete sweeps are shown, separated by the vertical grey dashed line.
The optimal rank-one approximation obtained from the SVD is
indicated for comparison.
	}
	\label{fig:bp_conv}
\end{figure}

In this reduced network, the environment of site \(\ell\) is then approximated from the two converged messages entering \(E_{\mathcal N}^{(\ell)}\) and a norm factor $g^{(\ell)}$, see Fig.~\ref{fig:bp_norm_simple}:
\begin{equation}
\begin{gathered}
\mathcal N_{\rm BP}^{(\ell)}
= g^{(\ell)}
\mu_{\ell\leftarrow \ell+1}
\otimes
\mu_{\ell-1\to \ell},
\quad
g^{(\ell)}
= \prod_{k\neq\ell} \tilde E_{\mathcal N}^{(k)}
\\[-0.5em]
\hspace{-2.5em}
\includegraphics[scale=0.7]{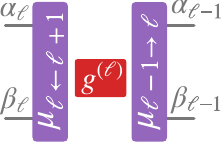}
\end{gathered}
\label{eq:bp_rank1}
\end{equation}
Therefore, BP messages provide a rank-one approximation of the environment norms $\mathcal N_{\rm BP}^{(\ell)}$ at every site $\ell$ of the ring. Note that this approximation can differ from the optimal rank-one approximation obtained from the leading singular triplet of each \(\mathcal N^{(\ell)}\) as in Eq.~\eqref{eq:svd_n}. Fig.~\ref{fig:bp_conv} shows the relative error norm 
\begin{equation}
\varepsilon_\mathcal{N}=\left|\mathcal{N}^{(1)}-\mathcal{N}^{(1)}_{\mathrm{BP}}\right|/\left|\mathcal{N}^{(1)}\right|
\end{equation} during BP-updates for a random ring of size $N=20$ and $\chi=10$. It converges toward a fix point close to the SVD rank-$1$ approximation.

\begin{figure}[h!!!!!!]
	\centering
	\includegraphics[width=0.49\textwidth]{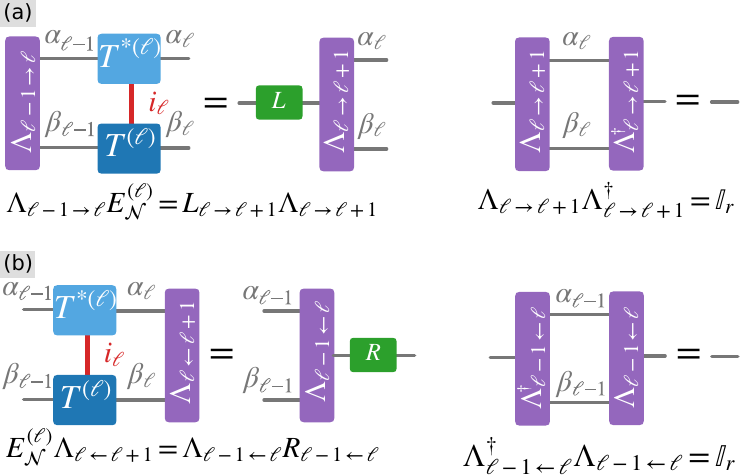}
	\caption{
		Higher rank Belief-propagation on a ring. Update of the clockwise (a) and counter-clockwise (b) messages.
	}
	\label{fig:bp_enlarged_messages}
\end{figure}

The rSVD suggests a complementary extension of BP. Inspired by the probe
matrix $\Omega$ in the rSVD procedure, we propagate matrix-valued messages
containing $r$ independent directions instead of a single vector message. Here, the clockwise messages are represented as matrices
$\Lambda_{\ell\to\ell+1}$  of size  $r\times \chi^2$, while the
counter-clockwise messages are represented as
$\Lambda_{\ell\leftarrow\ell+1}$ of size $\chi^2\times r$. After each update, the messages are orthogonalized (LQ decomposition). For the
clockwise propagation we perform (see Fig.~\ref{fig:bp_enlarged_messages})
\begin{equation}
\Lambda_{\ell-1\to\ell}
E_{\mathcal N}^{(\ell)}
=L_{\ell\to\ell+1}
\Lambda_{\ell\to\ell+1},
\label{eq:bp_big_lq}
\end{equation}
where $\Lambda_{\ell\to\ell+1}$ has orthonormal rows, $
\Lambda_{\ell\to\ell+1}
\Lambda_{\ell\to\ell+1}^{\dagger}
=
\mathbb I_r.
$
Equivalently, the counter-clockwise messages are chosen to have orthonormal
columns,
$
\Lambda_{\ell\leftarrow\ell+1}^{\dagger}
\Lambda_{\ell\leftarrow\ell+1}
=
\mathbb I_r.
$
After each counter-clockwise update, we perform the corresponding QR
decomposition,
\begin{equation}
E_{\mathcal N}^{(\ell)}\Lambda_{\ell\leftarrow\ell+1}
=
\Lambda_{\ell-1\leftarrow\ell}\,R_{\ell-1\leftarrow\ell}.
\label{eq:bp_big_qr}
\end{equation}
After convergence,
$\Lambda_{\ell-1\to\ell}$ and
$\Lambda_{\ell\leftarrow\ell+1}$ define $r$-dimensional
reduced subspaces of the left and right virtual spaces of
$\mathcal M^{(\ell)}$, respectively. 

\begin{figure}[h!]
	\centering
	\includegraphics[width=0.42\textwidth]{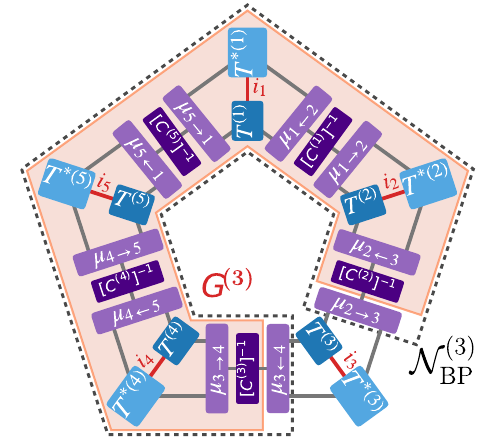}
	\caption{
		Norm of the ring state $\langle\phi|\phi\rangle$ evaluated at the higher rank BP fixed point. The conventions are the same as in Fig.~\ref{fig:bp_norm_simple}.
	}
	\label{fig:bp_norm_gen}
\end{figure}

Similarly to Eq.~\eqref{eq:bp_proj} we can insert between every pair of sites $\ell$ and $\ell+1$ the projector:
\begin{equation}
\begin{gathered}
P^{(\ell)}
=
\Lambda_{\ell\leftarrow\ell+1}
\underbrace{
	\left[
	\Lambda_{\ell\rightarrow\ell+1}
	\Lambda_{\ell\leftarrow\ell+1}
	\right]
}_{C^{(\ell)}}\mathord{}^{-1}
\Lambda_{\ell\rightarrow\ell+1}
\\[0.5em]
\hspace{1.5em}
\includegraphics[scale=0.7]{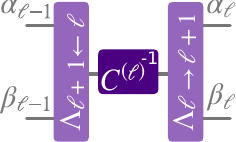}
\end{gathered}
\label{eq:bp_generalized_proj}
\end{equation}
Here $C^{(\ell)}$ of size $r\times r$ is the overlap matrix between the clockwise and counterclockwise message subspaces. Provided \(C^{(\ell)}\) is invertible, inserting \([C^{(\ell)}]^{-1}\) establishes a biorthogonal pairing between these bases, ensuring that \(P^{(\ell)}\) is a rank-\(r\) oblique projector. Similarly to Eq.~\eqref{eq:bp_reduced} it defines reduced \(r\times r\) transfer matrices as:
\begin{equation}
\begin{gathered}
\tilde E_{\mathcal N}^{(\ell)}
=
\Lambda_{\ell-1\to\ell}
E_{\mathcal N}^{(\ell)}
\Lambda_{\ell\leftarrow\ell+1}
\\[0.5em]
\hspace{3em}
\includegraphics[scale=0.7]{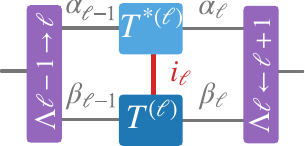}
\end{gathered}
\label{eq:bp_reduced}
\end{equation}
The norm $\braket{\phi|\phi}$ is then approximated by (see Fig.~\ref{fig:bp_norm_gen}):
\begin{equation}
\braket{\phi|\phi}
\approx\mathrm{Tr}\left(\prod_{k=1}^{N}[C^{(k)}]^{-1}
\tilde E_{\mathcal N}^{(k+1)}\right)
\end{equation}
A natural BP approximation of $\mathcal N^{(\ell)}$ can be
obtained by projecting $\mathcal N^{(\ell)}$ onto the subspaces generated by the messages:
\begin{equation}
\mathcal N_{\rm BP,proj.}^{(\ell)}
=
\Lambda_{\ell\leftarrow\ell+1}
\Lambda_{\ell\leftarrow\ell+1}^\dagger
\mathcal N^{(\ell)}\,
\Lambda_{\ell-1\to\ell}^\dagger
\Lambda_{\ell-1\to\ell}\,
.
\label{eq:generalized_bp_n}
\end{equation}
Equation~\eqref{eq:generalized_bp_n} requires contracting with the full environment
$\mathcal N^{(\ell)}$ at each site. Alternatively, a rank-$r$ approximation of $\mathcal N^{(\ell)}$ for all $\ell$ can be constructed directly from
the converged BP messages. This representation is particularly
useful for evaluating multiple observables or correlation
functions using the same compressed network. Inserting the projectors
of Eq.~\eqref{eq:bp_generalized_proj} on every bond yields the
reduced norm network of Eq.~\eqref{eq:bp_reduced}, from which
$\mathcal N^{(\ell)}$ can be approximated by contractions
around the ring (see Fig.~\ref{fig:bp_norm_gen}):
\begin{equation}
\begin{aligned}
\mathcal N_{\rm BP}^{(\ell)}
={}
\Lambda_{\ell\leftarrow\ell+1}
&[C^{(\ell)}]^{-1}
\tilde E_{\mathcal N}^{(\ell+1)}
[C^{(\ell+1)}]^{-1}\\&
\cdots
\tilde E_{\mathcal N}^{(\ell-1)}
[C^{(\ell-1)}]^{-1}
\Lambda_{\ell-1\to\ell}.
\label{eq:gbp}
\end{aligned}
\end{equation}
This can be expressed using the accumulated $L$/$R$ factors during the iterations. Indeed, using the message
update equations \eqref{eq:bp_big_lq}-\eqref{eq:bp_big_qr}, the reduced transfer matrices satisfy
\begin{equation}
\tilde E_{\mathcal N}^{(k)}
=
L_{k\to k+1}C^{(k)}
=
C^{(k-1)}R_{k-1\leftarrow k}.
\end{equation}
Denoting $R^{(\ell)}$ and $L^{(\ell)}$ the products around the ring:
\begin{align}
L^{(\ell)} &= L_{\ell+1\to \ell+2} \dots L_{\ell-1\to \ell},\\
R^{(\ell)} &= R_{\ell\leftarrow\ell+1}\dots R_{\ell-2\leftarrow\ell-1},
\end{align}
the environment approximation can be written as
\begin{equation}
\begin{gathered}
\mathcal N_{\rm BP}^{(\ell)}
=
\Lambda_{\ell\leftarrow\ell+1}
G^{(\ell)}
\Lambda_{\ell-1\to\ell}
\\[0.5em]
\hspace{3.5em}
\includegraphics[scale=0.7]{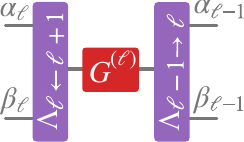}
\end{gathered}
\label{eq:generalized_bp_interpolatory}
\end{equation}

where
\begin{equation}
G^{(\ell)}
=
[C^{(\ell)}]^{-1}L^{(\ell)}
=
R^{(\ell)}[C^{(\ell-1)}]^{-1}.
\label{eq:R-C-L}
\end{equation}
Therefore, retaining the accumulated products \(R^{(\ell)}\) during BP gives the approximation of the effective norm at every site. 

Another interesting property of this construction is that it
interpolates the action of the exact environment $\mathcal N ^ {(\ell)}$ on both
converged message subspace. Iterating equation \eqref{eq:bp_big_lq}-\eqref{eq:bp_big_qr} around the ring gives:
\begin{align}
\Lambda_{\ell\leftarrow\ell+1}R^{(\ell)}
&=
\mathcal N^{(\ell)}
\Lambda_{\ell-1\leftarrow\ell},
\\
L^{(\ell)}\Lambda_{\ell-1\to\ell}
&=
\Lambda_{\ell\to\ell+1}
\mathcal N^{(\ell)}.
\end{align}
From Eq.~\eqref{eq:generalized_bp_interpolatory}-\eqref{eq:R-C-L} it follows immediately that
\begin{align}
\mathcal N_{\rm BP}^{(\ell)}
\Lambda_{\ell-1\leftarrow\ell}
&=
\mathcal N^{(\ell)}
\Lambda_{\ell-1\leftarrow\ell},
\\
\Lambda_{\ell\to\ell+1}
\mathcal N_{\rm BP}^{(\ell)}
&=
\Lambda_{\ell\to\ell+1}
\mathcal N^{(\ell)}.
\end{align}
The resulting approximation is therefore exact on the incoming and outgoing messages (in the spirit of cross interpolation \cite{Goreinov97}). 
\begin{figure}
	\centering
	\includegraphics[width=0.45\textwidth]{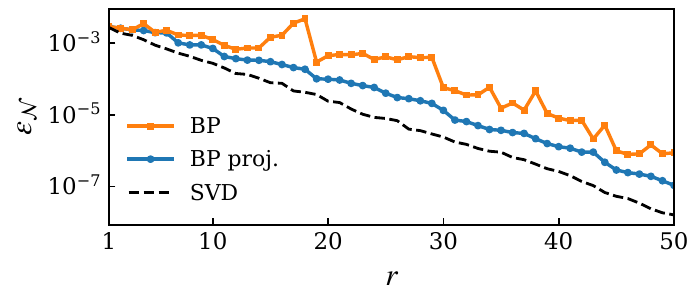}
	\caption{
		Generalized higher rank belief propagation compared to SVD. Relative norm error of  $\epsilon_{\mathcal N}$ at site $\ell=1$ of the ground state of the periodic spin-$1/2$ Hamiltonian with $N=100$ and $\chi=100$, shown as a function of the retained rank $r$. 
		The direct projected construction (BP proj.), is compared with the interpolatory BP construction (BP), and with the optimal rank-$r$ approximation given by the truncated SVD.
	}
	\label{fig:rbp}
\end{figure}

Figure~\ref{fig:rbp} compares the relative error $\varepsilon_\mathcal{N}$ of the effective norm $\mathcal{N}^{(1)}$ for the higher rank BP projected of Eq.~\eqref{eq:generalized_bp_n}, the interpolated version of Eq.~\eqref{eq:generalized_bp_interpolatory} and truncated SVD as a function of the retained rank $r$. We use the ground state of the periodic spin-$1/2$ Hamiltonian with $N=100$ and $\chi=100$. Although this higher rank BP does not achieve the same optimal accuracy as the truncated SVD, increasing the message rank gradually improves the approximation and clearly outperforms standard BP ($r=1$). The interpolated version remains noisy, but follows the same tendency.

This higher-rank BP construction may be particularly useful for extending loop expansions to two-dimensional tensor networks, such as PEPS. Defining the complementary projectors $Q^{(\ell)}=I-P^{(\ell)}$, the converged message equations \eqref{eq:bp_big_lq}–\eqref{eq:bp_big_qr} imply
\begin{equation}
Q^{(\ell-1)}E_{\mathcal N}^{(\ell)}P^{(\ell)}
=
P^{(\ell-1)}E_{\mathcal N}^{(\ell)}Q^{(\ell)}=0.
\end{equation}
This decoupling of retained and complementary subspaces provides a natural basis for systematic loop corrections beyond BP in higher-dimensional tensor networks \cite{Evenbly26,Midha26b}. The framework may also be extended to approximate environment contractions in PEPS and combined with block belief propagation \cite{Guo23,Tindall25}. Related recent work has explored tensor-train representations of messages, yielding matrix-product valued messages for PEPS \cite{Woolls26}.

\section{Conclusion}
\label{sec:conclusion}

In this work, we have developed a more robust and efficient formulation of Ring-DMRG based on low-rank periodic environments. Building on earlier compression and windowing strategies \cite{Pippan10,Rossini11}, we introduced a generalized Davidson solver and a balanced frame that improve the conditioning and stability of the local generalized eigenvalue problem while retaining matrix-free operations with leading cost $\mathcal{O}(p\chi^3)$.

Using the same compressed environments, we formulated a controlled two-site update for tensor rings. Unlike standard open-boundary DMRG, truncation in a ring is governed by a nontrivial norm metric and is therefore a weighted low-rank approximation problem. We showed that an appropriate change of frame can reduce this problem to an ordinary SVD when the metric is sufficiently separable, while a full-metric alternating least-squares scheme provides a more accurate treatment when several environment modes contribute significantly.

We identified the regimes in which Ring-DMRG can
be advantageous over tensor-train approaches. These
are characterized by a large ring-to-train bond-dimension overhead and a strongly compressible environment, as occurs naturally for sufficiently short correlation lengths and, more subtly, for large critical systems in the finite-size–finite-entanglement regime.

The low-rank environment picture also connects Ring-DMRG with belief propagation: ordinary BP corresponds to retaining a single dominant transfer direction, whereas the present construction retains a finite-dimensional subspace of long-lived modes. This suggests possible extensions toward systematic corrections to belief propagation on loopy tensor networks.

To become competitive in wall-clock performance, Ring-DMRG will require the modern implementation techniques used in state-of-the-art tensor-network libraries, including optimized contractions, block-sparse operations, and efficient memory management \cite{Fishman22}. Its more compact
representation also lowers memory requirements, which may be particularly advantageous for GPU-based implementations, where available memory can be a more severe limitation than the cost of the underlying linear algebra.
 It would also be interesting to combine low-rank periodic environments with tensor-cross and quantics methods \cite{NunezFernandez22,NunezFernandez2025,Jeannin25,Rohshap25}, or with alternative tensor-network constructions such as replica tensor trains \cite{Srdinsek26}. Such developments may extend the present framework to broader classes of periodic and higher-dimensional problems.

\begin{acknowledgments}
The authors thank Anna Kauch and Markus Wallerberger for helpful discussions and feedback on this work. This work was funded in part by the Austrian Science Fund (FWF) through Grant DOI 10.55776/P36332. We also acknowledge computational resources provided by the Arnold
Sommerfeld Center for theoretical physics
(www.theorie.physik.unimuenchen.de). This work was supported in part by
the Deutsche Forschungsgemeinschaft under Germany’s Excellence Strategy
EXC-2111 (Project No. 390814868). It is part of the Munich Quantum
Valley, supported by the Bavarian state government with funds
from the Hightech Agenda Bayern Plus.

\end{acknowledgments}

\appendix

\section{Building MPOs in ring form}
\label{app:ring_mpo}

We briefly describe how the periodic matrix product operators used in
this work are constructed. A ring MPO is written as
\begin{equation}
H(i_1,j_1,\ldots,i_N,j_N)
=
\operatorname{Tr}\!\left(
W^{(1)}_{i_1,j_1}
W^{(2)}_{i_2,j_2}
\cdots
W^{(N)}_{i_N,j_N}
\right),
\end{equation}
where, for fixed physical indices, $W^{(\ell)}_{i_\ell,j_\ell}$ is a matrix acting
on the MPO virtual space. The trace closes the virtual indices around
the ring.

The MPO virtual index can be viewed as the state of a finite automaton.
A nonzero block
\begin{equation}
\left[W^{(\ell)}_{i_\ell,j_\ell}\right]_{\omega_{\ell-1},\omega_\ell}
=
O_{i_\ell,j_\ell}
\end{equation}
means that the automaton enters site $\ell$ in state $\omega_{\ell-1}$, places the
operator $O$ on that site, and exits in state $\omega_\ell$. Thus each
closed virtual path generates one operator string, and the trace sums
over all allowed closed paths.

For the nearest-neighbor Heisenberg Hamiltonian
\begin{equation}
H =
\sum_{\ell=1}^{N}
\left(
S^x_\ell S^x_{\ell+1}
+
S^y_\ell S^y_{\ell+1}
+
S^z_\ell S^z_{\ell+1}
\right),
\end{equation}
we use five virtual states: a finished state $1$, three intermediate
states $2,3,4$ indicating that an $S^x$, $S^y$, or $S^z$ operator has
been started, and a not-yet-started state $5$. With the ordering
$(1,2,3,4,5)$, the bulk tensor is
\begin{equation}
\begingroup
\setlength{\arraycolsep}{4pt}
\renewcommand{\arraystretch}{1.15}
W_{\rm bulk}
=
\begin{pmatrix}
I   & 0   & 0   & 0   & 0 \\
S^x & 0   & 0   & 0   & 0 \\
S^y & 0   & 0   & 0   & 0 \\
S^z & 0   & 0   & 0   & 0 \\
0   & S^x & S^y & S^z & I
\end{pmatrix}.
\endgroup
\label{eq:bulk_heisenberg_ring_mpo}
\end{equation}
Rows denote incoming virtual states and columns outgoing states. From
state $5$ the automaton may either remain in $5$, placing an identity,
or move to one of the intermediate states and place the first spin
operator. From states $2,3,4$ it must move to state $1$, placing the
matching second spin operator. Once in state $1$, it only propagates
identities. The distinction between states $5$ and $1$ ensures that each
closed path contains exactly one interaction term, rather than a product
of several terms.

In an open-boundary MPO, boundary vectors select the initial and final
automaton states. In a ring MPO there are no such vectors, so one chooses
an arbitrary cut and modifies the tensor at that site. For the
nearest-neighbor Heisenberg MPO we use
\begin{equation}
\begingroup
\setlength{\arraycolsep}{4pt}
\renewcommand{\arraystretch}{1.15}
W_{\rm cut}
=
\begin{pmatrix}
0 & S^x & S^y & S^z & I \\
0 & 0   & 0   & 0   & S^x \\
0 & 0   & 0   & 0   & S^y \\
0 & 0   & 0   & 0   & S^z \\
0 & 0   & 0   & 0   & 0
\end{pmatrix}.
\endgroup
\label{eq:cut_heisenberg_ring_mpo}
\end{equation}
The block $\left[W_{\rm cut}\right]_{1,5}=I$ resets the automaton from the finished
sector to the not-yet-started sector across the cut, allowing paths to
close. The blocks $\left[W_{\rm cut}\right]_{1,2}$, $\left[W_{\rm cut}\right]_{1,3}$, and
$\left[W_{\rm cut}\right]_{1,4}$ start interactions at the cut site, while
$\left[W_{\rm cut}\right]_{2,5}$, $\left[W_{\rm cut}\right]_{3,5}$, and $\left[W_{\rm cut}\right]_{4,5}$ finish
interactions that wrap from site $N$ to site $1$. The cut tensor breaks the manifest translational invariance of the MPO representation, but not of the represented operator. This is a standard preferable finite-size construction \cite{Weyrauch13}: enforcing exact site-independent tensors in a closed finite MPO would require increasing the bond dimension.
\section{Environment scaling}
\label{app:env_scaling}
\begin{figure}
	\centering	\includegraphics[width=0.48\textwidth]{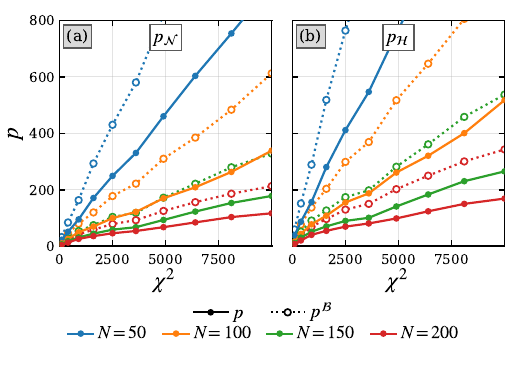}
	\caption{
		Scaling of the retained environment ranks with the tensor ring bond dimension for the Heisenberg chain, for a truncation $\epsilon_\text{SVD}=10^{-12}$. (a) Mean retained rank $p_{\mathcal N}$ of the norm environment and (b) mean retained rank $p_{\mathcal H}$ of the Hamiltonian environment, shown as functions of $\chi^2$ for $N=50$, $100$, $150$, and $200$. Solid lines show the ranks of the site-dependent environments, while dotted lines show the corresponding ranks $p_{\mathcal N}^{\mathcal B}$ and $p_{\mathcal H}^{\mathcal B}$ of the windowed environments, using $n_{\mathrm{open}}=2N/3$.
	}
	
	\label{fig:env_chi2}
\end{figure}

Figure~\ref{fig:env_chi2} shows that both environment ranks, \(p_{\mathcal N}\) and \(p_{\mathcal H}\), scale approximately linearly with \(\chi^2\). The windowed construction leaves this scaling essentially unchanged and increases the retained rank only moderately, by roughly a factor of two for \(n_{\mathrm{open}}=2N/3\).

\section{Implementation of U(1) symmetry}
\label{app:sym}

We also provide an implementation exploiting abelian $U(1)$ symmetries \cite{Singh07,Rakov16,Hauschild18},
such as the total magnetization $\sum_k S^z_k$ of the Heisenberg chain,
which restricts the optimization to a target sector and reduces both
memory and computational cost by working only on the nonzero blocks.
In this setting every index carries a charge: each physical state $i_k$ has
a definite charge $q(i_k)$ (for spin $1/2$, $q(\uparrow)=+1/2$ and
$q(\downarrow)=-1/2$), and each virtual index $\alpha_k$ is assigned a
charge $q(\alpha_k)$ labelling the total charge accumulated up to bond $k$.
The cores are then nonzero only for index combinations satisfying the
conservation rule
\begin{equation}
q(\alpha_{k-1})+q(i_k)=q(\alpha_k),
\label{eq:charge_rule}
\end{equation}
so that $T^{(k)}$ is block sparse and all local operations (contractions,
QR, SVD) are simply performed block by block.

The only nontrivial point is the low-rank environment. In the randomized
SVD of Sec.~\ref{sec:rsvd}, the probing matrix $\Omega$ cannot be dense,
as this would violate Eq.~\eqref{eq:charge_rule}: it must be drawn only in
the charge sectors shared by the incoming and outgoing doubled bond spaces,
so that the factors $U^p$ and $V^p$ of Eq.~\eqref{eq:svd_n} and \eqref{eq:svd_h} remain valid symmetric transfer
tensors. 

The changing of frame gauge matrices $G_U$, $G_V$ are then block
diagonal in the boundary charge, and their inverse square roots, as well as
the metric solves in Eq.~\eqref{eq:normal_ed}, are performed independently
in each sector. 

Finally, closing the ring imposes
$q(\alpha_N)=q(\alpha_0)$, i.e. $\sum_k q(i_k)=0$: the present
implementation is restricted to the charge-neutral sector.

When Abelian symmetries are enforced, a traditional one-site update cannot modify the charge sectors and their multiplicities carried by the virtual bonds. A two-site update is therefore particularly important, as it allows the symmetry-sector structure and the bond dimension to adapt dynamically during the optimization \cite{Gleis23}. We introduce such an update in Sec.~\ref{sec:two_site}, together with an environment-aware truncation procedure suitable for tensor rings.

\section{Sum of two tensor rings}
\label{App:ring_addition}
\begin{figure}[t]
	\centering
	\includegraphics[width=\columnwidth]{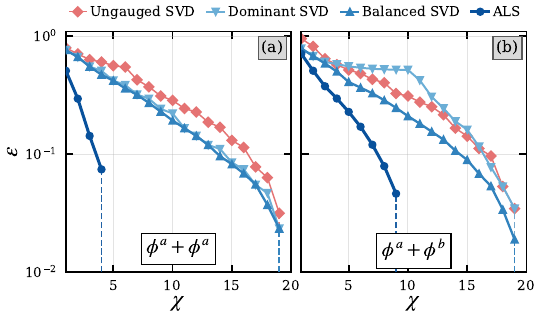}
	\caption{
		Two-site compression after exact addition of random tensor rings with $N=20$
		and initial bond dimension $\chi=10$ between two sites. The relative norm error $\epsilon$ of
		the compressed tensor ring is shown as a function of the retained bond
		dimension $\chi$. Panel (a) corresponds to the self-addition
		$\phi^a+\phi^a$, while panel (b) shows the sum $\phi^a+\phi^b$ of two
		independently generated rings. The vertical line indicates the bond
		dimension for which the compression becomes exact, $\epsilon=0$.
	}
	\label{fig:ring_addition}
\end{figure}

As for tensor trains, two tensor rings can be summed by block-diagonal
concatenation of their cores. The main difficulty is the subsequent
compression. Even self-addition illustrates the problem: adding a ring of bond
dimension $\chi$ to itself produces a tensor ring with bond
dimension $2\chi$ on every bond, while the tensor is simply $2\phi$
and therefore admits an exact representation with bond dimension $\chi$. Unlike for tensor trains, however, a sequence of local SVD truncations does not in general recover such a reduced tensor-ring representation \cite{Batselier18,Ye18}. Alternative constructions based on TT-like paths have been proposed \cite{Mickelin20,Ellwein26}. Here, the full-environment two-site ALS truncation introduced in Sec.~\ref{sec:two_site} provides a different route: instead of determining the reduced bond dimension from local information alone, it optimizes the two-site factorization in the metric defined by the complete tensor-ring environment.

To illustrate this, we consider two random tensor rings $\phi^a$ and $\phi^b$ with $N=20$ sites and bond dimension $\chi=10$. Their exact block-diagonal addition initially produces a tensor ring with bond dimension $2\chi=20$. Figure~\ref{fig:ring_addition} shows the truncation error obtained when compressing the bond between sites $0$ and $1$. For the self-addition $\phi^a+\phi^a$ [panel (a)], the ALS compression reaches exactly zero error for retained bond dimensions $\chi\geq5$. For the sum of two independently generated rings, $\phi^a+\phi^b$ [panel (b)], the error becomes exactly zero for $\chi\geq10$. Thus, the direct block-diagonal construction introduces redundant virtual dimensions that can be removed exactly by the full-environment ALS compression. Note that a sequence of two-site truncations may be suboptimal because each step only optimizes a single bond. A more global ALS compression can instead optimize the cores sequentially around the ring, keeping the updates accelerated by the compressed environments.

\section{Convert a ring into a train}
\label{app:ring_to_train}

We recall the elementary construction showing that any tensor ring can be
written exactly as a tensor train, at the price of increasing the bond
dimensions. The point is simply to cut one virtual bond of the ring and
propagate the corresponding index through the open chain.

Let $\phi$ be a tensor ring,
\begin{align}
\phi(i_1,\ldots,i_N)
=
\sum_{\{\alpha_\ell\}}
\left[T^{(1)}_{i_1}\right]_{\alpha_N,\alpha_1}
\cdots
\left[T^{(N)}_{i_N}\right]_{\alpha_{N-1},\alpha_N}.
\end{align}
We now cut the ring on the bond \(\alpha_N\).
The new tensor-train bond indices are then pairs
\begin{equation}
b_\ell=(\alpha_N,\alpha_\ell),
\qquad
\ell=1,\ldots,N-1,
\end{equation}
so that \(b_\ell\) labels the product space $
\{1,\ldots,\chi_N\}\times \{1,\ldots,\chi_\ell\}
$. Define the open-boundary tensor-train cores:
\begin{align}
\left[A^{(1)}_{i_1}\right]_{(\alpha_N,\alpha_1)}
&=
\left[T^{(1)}_{i_1}\right]_{\alpha_N,\alpha_1},\\
\left[A^{(\ell)}_{i_\ell}\right]_{(\alpha_N,\alpha_{\ell-1}),(\alpha_N',\alpha_\ell)}
&=
\delta_{\alpha_N,\alpha_N'}\,
\left[T^{(\ell)}_{i_\ell}\right]_{\alpha_{\ell-1},\alpha_\ell},
\label{eq:ring_to_train}\\
\left[A^{(N)}_{i_N}\right]_{(\alpha_N,\alpha_{N-1})}
&=
\left[T^{(N)}_{i_N}\right]_{\alpha_{N-1},\alpha_N}.
\end{align}
The Kronecker symbol in Eq.~\eqref{eq:ring_to_train} ensures that
the same cut index \(\alpha_N\) is transported through the whole train. The price of opening the ring is that each tensor-train bond carries both
the original ring bond index \(\alpha_\ell\) and the cut index \(\alpha_N\). Therefore,
for a cut on the bond of dimension \(\chi_N\), the new bond dimensions $\widetilde \chi_\ell$ verifies
\begin{equation}
\widetilde \chi_\ell \le \chi_N\chi_\ell,
\qquad \ell=1,\ldots,N-1.
\label{eq:ring_to_train_rank_bound}
\end{equation}
In the homogeneous case \(\chi_\ell=\chi\), this gives
\begin{equation}
\widetilde \chi \le \chi^2 .
\end{equation}

This construction is not necessarily rank optimal. After converting the
ring into a train, one can apply standard tensor-train rounding by QR and
SVD decompositions to reduce the ranks. Nevertheless,
Eq.~\eqref{eq:ring_to_train_rank_bound} gives the useful worst-case
bound: a tensor ring of bond dimension \(\chi\) can always be represented
exactly as a tensor train with bond dimension at most \(\chi^2\).
Conversely, this explains why representing periodic correlations with an
open-boundary tensor train may require a squared bond dimension.

\begin{figure}
	\centering
	\includegraphics[width=0.48\textwidth]{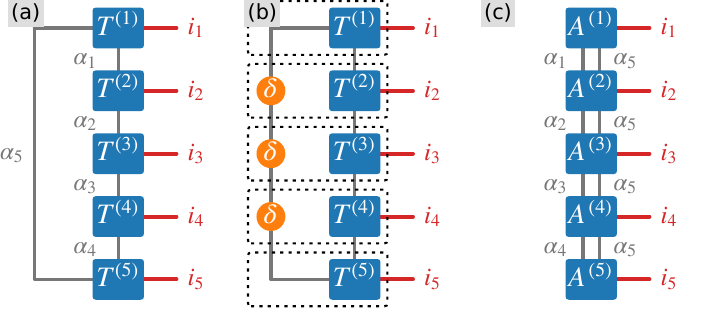}
	\caption{
		Exact mapping of a tensor ring onto a tensor train, illustrated for $N=5$.
		(a) Tensor ring: the closing bond $\alpha_N$ connects the first and last
		cores, encoding the periodic boundary conditions through the trace.
		(b) The closing bond is unfolded and carried along the chain as a
		spectator index, with Kronecker deltas $\delta_{\alpha_N\alpha_N'}$
		inserted at each intermediate site. Dotted boxes indicate the tensors
		that are grouped together at each site.
		(c) Absorbing the spectator wire into each core,
		$A^{(\ell)} = T^{(\ell)}\otimes\delta$, yields a tensor train in which each
		internal bond carries the pair $(\alpha_N,\alpha_\ell)$, i.e. an effective
		bond dimension $\chi^2$. 
	}
	\label{fig:ring_to_train}
\end{figure}

\section{Algorithmic cost}
\label{app:algorithmic_cost}

We summarize the leading cost of the one-site Ring-DMRG algorithm, assuming throughout the windowing scheme introduced above. The tensor ring requires \(\mathcal{O}(Nd\chi^2)\) memory. In addition, the contractions needed throughout a window of \(n_{\rm open}\) sites require \(\mathcal{O}(n_{\rm open}p_{\mathcal N}\chi^2)\) for the norm environment and \(\mathcal{O}(n_{\rm open}p_{\mathcal H}\chi_W\chi^2)\) for the Hamiltonian environment. Since we choose \(n_{\rm open}\propto N\), the total memory cost is \(\mathcal{O}[N\chi^2(d+p_{\mathcal N}+\chi_Wp_{\mathcal H})]\), up to the workspace of the iterative solver.

For each local optimization, applying the compressed norm and Hamiltonian environments costs \(\mathcal{O}(d p_{\mathcal N}\chi^3)\) and \(\mathcal{O}(d p_{\mathcal H}\chi_W\chi^3)\), respectively. Treating the number of Davidson iterations as a prefactor, and using \(n_{\rm open}\propto N\) so that environment construction is also linear in \(N\), the leading cost of a sweep is \(\mathcal{O}[Nd\chi^3(p_{\mathcal N}+\chi_Wp_{\mathcal H})]\), up to the prefactors associated with the iterative solver and randomized SVD.

For comparison, standard DMRG in tensor-train form requires \(\mathcal{O}(Nd\chi^2)\) memory and \(\mathcal{O}(Nd\chi^3)\) operations per sweep. However, as discussed in Sec.~\ref{app:ring_to_train}, reaching the same accuracy can require a substantially larger bond dimension in tensor-train form than in tensor-ring form. This reduction in bond dimension may therefore compensate for the additional cost of the periodic environments. In Sec.~\ref{sec:regime}, we study quantitatively when this trade-off leads to a computational advantage for Ring-DMRG.

\begin{figure}[t]
	\centering
	\includegraphics[width=\columnwidth]{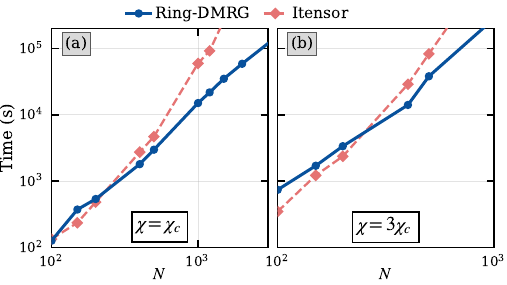}
	\caption{
		Comparison of the estimated convergence times of Ring-DMRG (solid blue) and tensor-train DMRG implemented using ITensor (dashed red) for the periodic spin-$1/2$ Heisenberg chain. We consider two bond-dimension trajectories: (a) $\chi_r=\chi_c(N)$ and (b) $\chi_r=3\chi_c(N)$, with the ITensor bond dimension chosen as $\chi_t=\chi_r^{1.5}$.
	}
	\label{fig}
\end{figure}

In Fig.~\ref{fig}, we compare the convergence times of Ring-DMRG and conventional tensor-train DMRG implemented using ITensor~\cite{Fishman22} for the periodic spin-$1/2$ Heisenberg chain. We follow the two trajectories $\chi_r=\chi_c(N)$ and $\chi_r=3\chi_c(N)$ introduced in Sec.~\ref{sec:regime}, choosing the ITensor bond dimension according to $\chi_t=\chi_r^{1.5}$ (see Fig.~\ref{fig:N500}). In both regimes, Ring-DMRG exhibits an increasing computational advantage with system size.

\section{A low-rank environment with a quadratic ring-to-train overhead}
\label{app:independent_rank_examples}

We give an explicit family of tensor rings showing that the rank of a
long norm environment and the cost of opening the ring into a tensor
train can scale independently. Within this construction, the exact
environment rank can be varied from $p_{\mathcal N}=2$ up to the maximal
value $p_{\mathcal N}=\chi_{\rm r}^2$, while the tensor-train bond
dimension remains $\chi_{\rm t}=\chi_{\rm r}^2$.

Let the virtual space of each bond be
$\mathbb C^\chi=(\mathbb C^2)^{\otimes m}$, so that $\chi=2^m$.
We take \(N=2m\) sites with physical dimension \(d=4\) identified with the four states $(P_1,P_2,P_3,P_4)=(I,X,Y,Z)$. The ring is composed of \(m\) different types of cores repeated twice,
$(1,\ldots,m,1,\ldots,m)$. A site \(\ell\) is associated with the virtual qubit
\begin{equation}
q(\ell)=
\begin{cases}
\ell, & 1\leq \ell\leq m,\\
\ell-m, & m<\ell\leq 2m.
\end{cases}
\end{equation}
Thus sites \(q\) and \(m+q\) are the two cores of type \(q\), and are
chosen identically. For a core \(\ell\) of type \(q\), the four physical slices
are
\begin{equation}
T^{(\ell)}_{i_\ell}
=
\sqrt{p_{i_\ell}^{[q]}}\,
I^{\otimes(q-1)}\otimes P_{i_\ell}
\otimes I^{\otimes(m-q)}.
\label{eq:example_core}
\end{equation}
where \(P_{i_\ell}\) denotes the Pauli matrix associated with the physical
index \(i_\ell\), and \(p_{i_\ell}^{[q]}\) is a weight associated with this
slice. Equivalently, writing the virtual indices of
$(\mathbb C^2)^{\otimes m}$ as binary strings
$\alpha_{\ell-1}=(a_1,\ldots,a_m)$ and
$\alpha_\ell=(b_1,\ldots,b_m)$, one has
\begin{equation}
\left[T^{(\ell)}_{i_\ell}\right]_{\alpha_{\ell-1},\alpha_\ell}
=
\sqrt{p_{i_\ell}^{[q]}}\,
(P_{i_\ell})_{a_qb_q}
\prod_{r\neq q}\delta_{a_rb_r}.
\label{eq:example_core_components}
\end{equation}
Hence each core acts nontrivially on only one of the \(m\) factors
forming the virtual bond.

For a core of type \(q\), the norm transfer operator \(E_{\mathcal N}^{(\ell)}\) defined in Eq.~\eqref{eq:transfer_N} acts on the doubled virtual space
$$
\mathcal V\otimes\overline{\mathcal V}
\simeq
\operatorname{End}(\mathcal V),
\qquad
\mathcal V=(\mathbb C^2)^{\otimes m},
$$
which has dimension \(\chi^2=(2^m)^2\). Since the local tensor in Eq.~\eqref{eq:example_core} acts nontrivially only on the \(q\)-th virtual qubit, the corresponding norm transfer map factorizes as
\begin{equation}
E_{\mathcal N}^{(\ell)}
=\widetilde I^{\otimes(q-1)}
\otimes
\widetilde E^{[q]}
\otimes
\widetilde I^{\otimes(m-q)},
\end{equation}
where \(\widetilde I\) denotes the identity superoperator on \(\operatorname{End}(\mathbb C^2)\). The nontrivial one-qubit transfer map
\(\widetilde E^{[q]}:\operatorname{End}(\mathbb C^2)\to\operatorname{End}(\mathbb C^2)\)
acts on a \(2\times2\) operator \(A\) as
\begin{equation}
\widetilde E^{[q]}(A)
=\sum_{i=1}^{4}
p_i^{[q]} P_i A P_i^\dagger .
\label{eq:example_pauli_channel}
\end{equation}
The Pauli basis diagonalizes $\widetilde E^{[q]}$. Explicitly,
\begin{align}
\widetilde E^{[q]}(P_1)
&=
\left(
p_1^{[q]}+p_2^{[q]}+p_3^{[q]}+p_4^{[q]}
\right) P_1
=
P_1,
\label{eq:example_pauli_eigenvalues_1}
\\
\widetilde E^{[q]}(P_2)
&=
\left(
p_1^{[q]}+p_2^{[q]}-p_3^{[q]}-p_4^{[q]}
\right) P_2,
\label{eq:example_pauli_eigenvalues_2}
\\
\widetilde E^{[q]}(P_3)
&=
\left(
p_1^{[q]}-p_2^{[q]}+p_3^{[q]}-p_4^{[q]}
\right) P_3,
\label{eq:example_pauli_eigenvalues_3}
\\
\widetilde E^{[q]}(P_4)
&=
\left(
p_1^{[q]}-p_2^{[q]}-p_3^{[q]}+p_4^{[q]}
\right) P_4.
\label{eq:example_pauli_eigenvalues_4}
\end{align}

When computing $\mathcal N^{(\ell)}$ (or $\mathcal N^{(\ell,\ell+1)}$), site $\ell$ (or sites $\ell$ and $\ell+1$) is removed. Since each core type is repeated twice around the ring, the remaining contraction still factorizes over the $m$ virtual qubits. If the transfer map associated with virtual qubit $q$ has $r_q$ nonzero Pauli modes, the full environment therefore contains all tensor products of these surviving modes. Its exact rank is thus
\begin{equation}
p_{\mathcal N}
=\prod_{q=1}^{m} r_q.
\end{equation}
By varying the weights $p_i^{[q]}$, the number of nonzero Pauli modes
of each one-qubit transfer map can be chosen to be
\begin{equation}
r_q\in\{1,2,3,4\}.
\end{equation}

We now show that, independently of the chosen environment rank
$p_{\mathcal N}$, the same family requires a tensor-train bond dimension
$\chi_{\rm TT}=\chi^2$, provided that all four weights $p_i^{[q]}$ are
strictly nonzero. For this purpose, consider the cut between the two
repetitions,
\begin{equation}
(1,2,\ldots,m)
\;\big|\;
(1,2,\ldots,m).
\nonumber
\end{equation}
The first occurrence of every virtual qubit $q$ lies on the left of the
cut, while its second occurrence lies on the right.

To determine the TT bond dimension across this cut, we reshape the full
tensor into a matrix
\begin{equation}
M\bigl[
(i_1,\ldots,i_m),
(i_{m+1},\ldots,i_{2m})
\bigr]
=
\phi(i_1,\ldots,i_{2m}).
\end{equation}
The rank of this matrix is precisely the Schmidt rank across the cut and
therefore gives the minimal TT bond dimension required at this bond.

Recall that a core of type $q$ acts only on the $q$-th factor of the
virtual space. Operators associated with different $q$'s therefore act
on different tensor-product factors and commute. The trace defining the
tensor ring can consequently be regrouped independently for each virtual
qubit. For a given virtual qubit $q$, the corresponding matrix is
\begin{equation}
M_q(i,j)
=
\sqrt{p_i^{[q]}p_j^{[q]}}\,
\operatorname{Tr}(P_iP_j).
\label{eq:example_local_flattening}
\end{equation}
Thus, up to a permutation of rows and columns, the full flattening is
the Kronecker product
\begin{equation}
M
=
M_1\otimes M_2\otimes\cdots\otimes M_m.
\label{eq:example_kron_flattening}
\end{equation}

Using the orthogonality of the Pauli matrices,
$\operatorname{Tr}(P_iP_j)=2\delta_{ij}$, one obtains
\begin{equation}
M_q(i,j)
=
2p_i^{[q]}\delta_{ij}.
\end{equation}
The weights used to tune the environment rank can be chosen such that all
four $p_i^{[q]}$ remain strictly positive. Therefore, independently of
the rank $r_q$ of the corresponding transfer map, $M_q$ is a diagonal
$4\times4$ matrix with four nonzero diagonal entries, and hence
\begin{equation}
\operatorname{rank}(M_q)=4
\end{equation}
for every $q$. The rank of the full flattening is therefore
\begin{equation}
\operatorname{rank}(M)
=
\prod_{q=1}^{m}\operatorname{rank}(M_q)
=
4^m.
\end{equation}
Since $\chi=2^m$,
\begin{equation}
\operatorname{rank}(M)
=
4^m
=
(2^m)^2
=
\chi^2.
\end{equation}
Any exact tensor train representing the same tensor must therefore have
a bond dimension of at least $\chi^2$ across this cut. Since opening a
tensor ring of bond dimension $\chi$ gives the general upper bound
$\chi_{\rm TT}\leq\chi^2$, this bound is saturated,
\begin{equation}
\chi_{\rm t}
=
\chi_{\rm r}^2.
\label{eq:example_tt_rank}
\end{equation}

The quadratic ring-to-train overhead is therefore independent of the
environment rank within this family. Varying the weights $p_i^{[q]}$
changes the number of surviving Pauli modes of the norm environment,
and hence $p_{\mathcal N}$, while keeping all four weights strictly
positive preserves the full rank of each local flattening $M_q$. Each
of the $m$ virtual factors therefore contributes four independent Schmidt
directions across the cut, giving a total of $4^m=\chi^2$ independent
directions.

\begin{figure}
	\centering
	\includegraphics[width=0.48\textwidth]{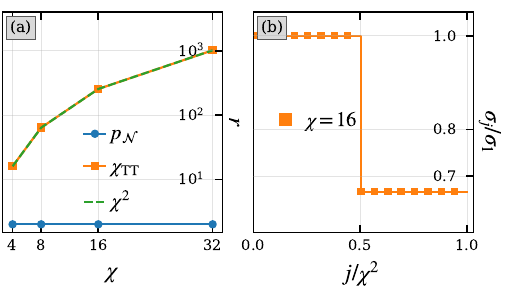}
	\caption{
		Explicit tensor-ring family with a low-rank norm environment but a
		quadratic ring-to-train overhead.
		(a) Environment rank \(p_{\mathcal N}\), tensor-ring bond dimension
		\(\chi\), and tensor-train bond dimension \(\chi_{\rm TT}\) obtained
		after exact ring-to-train conversion followed by QR--SVD rounding with
		relative tolerance \(10^{-10}\). The environment rank remains fixed at
		\(p_{\mathcal N}=2\), while the tensor-train bond dimension saturates
		the quadratic bound \(\chi_{\rm TT}=\chi^2\).
		(b) Normalized singular values of the flattening across the cut between
		the two repetitions. For the weights
		\(p_1=p_2=3/10\) and \(p_3=p_4=1/5\), the spectrum consists of two
		finite plateaus with ratio \(2/3\), showing that the
		\(\chi^2\) Schmidt directions are not associated with numerically small
		singular values and therefore cannot be removed by standard TT
		rounding.
	}
	\label{fig:env_singular_values_t}
\end{figure}

For the example shown in Fig.~\ref{fig:env_singular_values_t}, we choose
the particularly simple case $p_{\mathcal N}=2$. We set
\begin{equation}
p_1^{[q]}
=
p_2^{[q]}
=
p_3^{[q]}
=
p_4^{[q]}
=
\frac{1}{4},
\qquad q>1,
\end{equation}
so that the corresponding transfer maps have rank one and only the
identity Pauli mode survives. For the distinguished virtual qubit
$q=1$, we choose
\begin{equation}
p_1^{[1]}=p_2^{[1]}=\frac{3}{10},
\qquad
p_3^{[1]}=p_4^{[1]}=\frac{1}{5},
\end{equation}
for which the transfer map has the two nonzero modes
\begin{equation}
P_1\mapsto P_1,
\qquad
P_2\mapsto \frac{1}{5}P_2.
\end{equation}
Hence $r_1=2$ and $r_q=1$ for $q>1$, giving
\begin{equation}
p_{\mathcal N}
=
2\times 1^{m-1}
=
2.
\end{equation}

\section{Singular values of a product of transfer matrices}
\label{app:tranfer_prod}
We examine the scaling of the singular values of products of norm transfer matrices with system size $N$ and bond dimension $\chi$. Figure~\ref{fig:tranfer_matrix}(a) shows that the ratio of the two leading singular values of a product of $L$ consecutive transfer matrices (see Eq.~\eqref{eq:prod_transfer}) decays approximately exponentially with $L$, allowing us to extract an effective decay length $\xi$. As shown in Fig.~\ref{fig:tranfer_matrix}(b), the resulting values are well described by the phenomenological scaling form $N/\xi = b_1+b_2 N/\xi_\chi$ introduced in Eq.~\eqref{eq:decay_fss}, supporting the combined contribution of finite-size and finite-entanglement effects.

\begin{figure}
	\centering
	\includegraphics[width=\linewidth]{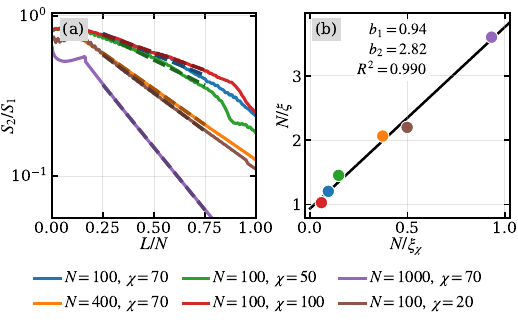}
	\caption{ 
		Scaling of transfer-environment singular values. (a) Ratio of the two leading singular values, $S_2^{(\ell)}(L)/S_1^{(\ell)}(L)$, of a product of $L$ consecutive transfer matrices starting at site 1, plotted as a function of the normalized product length $L/N$ for ground-state tensor rings with different system sizes $N$ and bond dimensions $\chi$ in the spin-$1/2$ Heisenberg model. Dashed lines indicate exponential fits. (b) Ratio $N/\xi$ plotted against $N/\xi_\chi$. The solid line represents a linear fit to $N/\xi=b_1+b_2N/\xi_\chi$, yielding $b_1=0.937$ and $b_2=2.822$, with $R^2=0.990$.
	}
	\label{fig:tranfer_matrix}
\end{figure}

\section{$J_1$-$J_2$ convergence}
\label{App:J1J2}
\begin{figure}[h!]
	\centering
	\includegraphics[width=0.48\textwidth]{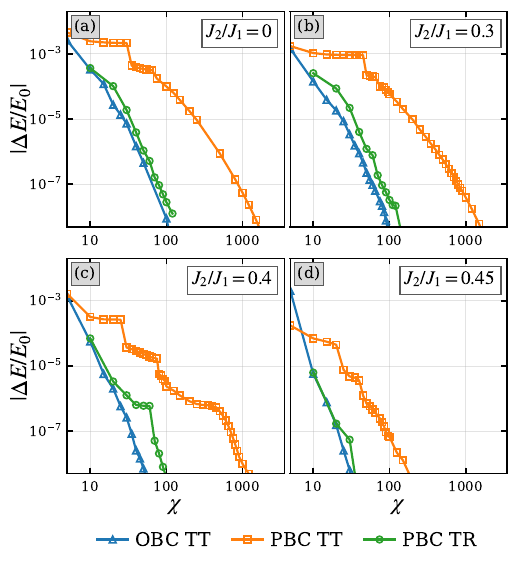}
	\caption{
		Convergence of the ground-state energy with bond dimension $\chi$ for the spin-$1/2$ $J_1$--$J_2$ Heisenberg chain with $N=200$, for (a) $J_2/J_1=0$, (b) $0.30$, (c) $0.40$, and (d) $0.45$. We compare standard DMRG in tensor-train form with open boundary conditions (OBC TT), standard DMRG in tensor-train form with periodic boundary conditions (PBC TT), and Ring-DMRG with a tensor-ring ansatz (PBC TR). For OBC TT, the relative energy error is measured with respect to the converged OBC energy. For PBC TT and PBC TR, the error is measured with respect to the converged PBC TT energy. Across all four values of $J_2/J_1$, the tensor-ring representation reaches a given energy accuracy at substantially smaller bond dimension than the periodic tensor-train representation.
	}
	
	\label{fig:J1J2-data}
\end{figure}

Figure~\ref{fig:J1J2-data} compares the bond-dimension convergence for the spin-$1/2$ \(J_1\)--\(J_2\) Heisenberg chain at \(J_2/J_1=0,\,0.30,\,0.40,\) and \(0.45\). In all four cases, the periodic tensor ring reaches a given energy accuracy at a substantially smaller bond dimension than the periodic tensor train, with $\chi_{\mathrm{TT}} \approx \chi_{\mathrm{TR}}^{1.5}.$ The required bond dimensions also decrease as \(J_2/J_1\) approaches the Majumdar--Ghosh point at \(J_2/J_1=1/2\).

\bibliographystyle{apsrev4-2}
\bibliography{references}

\end{document}